\documentclass{aa}  

\usepackage{graphicx}
\usepackage{txfonts}
\usepackage{lipsum}
\usepackage{subcaption}         
\usepackage{lscape}             
\usepackage{placeins}           
\usepackage{hyperref}
\usepackage{xspace}
\usepackage{tabularx}
\usepackage{multirow}
\usepackage{multicol}
\usepackage{xcolor}
\usepackage{booktabs}

\newcommand{\ngc}{NGC\,2359\xspace}
\newcommand{\kms}{\,km\,s$^{-1}$\xspace} 
\newcommand{\Msunyr}{\,$M_\odot$\,yr$^{-1}$\xspace} 

\begin{document}

   \title{Investigating particle acceleration in the Wolf-Rayet bubble \ngc}
   \author{Anindya Saha\inst{1,2,3 \thanks{Boya Fellow}}, 
         Anandmayee Tej\inst{1},  
         Santiago del Palacio\inst{4}, 
         Micha\"{e}l De Becker\inst{2},
         Paula Benaglia\inst{5},
         Ramananda Santra\inst{6},
         Ishwara Chandra CH\inst{6}
        }

   \institute{Indian Institute of Space Science and Technology, Thiruvananthapuram 695 547, Kerala, India\\
        \email{anindya.s1130@gmail.com}
        \and Space sciences, Technologies and Astrophysics Research (STAR) Institute, University of Li\`{e}ge, Belgium 
        \and Kavli Institute for Astronomy and Astrophysics, Peking University, Beijing 100871, People’s Republic of China 
        \and Department of Space, Earth and Environment, Chalmers University of Technology, Gothenburg, Sweden  
        \and Instituto Argentino de Radioastronomía (CONICET--CICPBA--UNLP), C.C No 5. 1894 Villa Elisa, Argentina
        \and National Centre for Radio Astrophysics, Pune 411 007, Maharashtra, India
        }    
   \date{Received Month date, year}

  \abstract
   { 
   Massive stars have been proposed as candidates to be major factories of Galactic cosmic rays (GCRs). However, this claim lacks enough empirical evidence, especially in the case of isolated stars. The powerful stellar winds from massive stars impact the ambient medium and produce strong shocks suitable for accelerating relativistic particles. The detection of non-thermal emission--particularly synchrotron emission in low-frequency radio bands--serves as a key proof of particle acceleration sites.}
   %
   %
   {We aim to assess the potential of isolated massive stars as sources of GCRs. }
   %
   %
   {We observed the Wolf-Rayet bubble, NGC\,2359, using the upgraded Giant Metrewave Radio Telescope at Band 3 (250--500~MHz) and Band 4 (550--950~MHz). Additionally, we utilized complementary archival radio datasets across different frequencies to derive the broad spectral energy distribution (SED) for several regions within the bubble. 
   In addition, to further characterize the interaction between the stellar wind and the ambient medium, we introduced a composite SED model including synchrotron and free--free emission, and two low-frequency turnover processes, the Razin-Tsytovich (RT) effect and free--free absorption (FFA). We used a Bayesian inference approach to fit the SEDs and constrain the electron number density and magnetic field strength.
   }
  %
   {The SEDs of several regions across the bubble reveal spectral indices steeper than $-0.5$, indicative of synchrotron radiation. Furthermore, the SEDs show a turnover below $\sim$1~GHz. 
   Our SED modelling suggests that the observed turnover is primarily caused by the RT effect, with a minor contribution from internal FFA.
   }
   %
   {Our analysis confirms the presence of synchrotron radiation within NGC\,2359. 
   This is the second detection of non-thermal emission in a stellar bubble surrounding a WR star, reinforcing the idea that such environments are sites of relativistic particle acceleration. This finding further supports the hypothesis that isolated massive stars are sources of GCRs of at least GeV energies.
   }


   \authorrunning{A. Saha et al.}
   \maketitle
\nolinenumbers

\section{Introduction}
\label{sec:intro}

Massive stars profoundly influence their surrounding interstellar medium (ISM) through mechanical, radiative and chemical feedback, particularly in their late evolutionary stages. 
In these phases, their intense radiation field drives powerful, supersonic stellar winds. 
As these winds propagate, they can interact either with the ambient ISM or, in the case of a binary system, collide with the wind from a companion star. 
This leads to two possible interaction scenarios: {\it wind--wind} and {\it wind--ISM} interactions, which generate high Mach number shocks ---suitable sites for relativistic particle acceleration through diffusive shock acceleration (DSA; e.g. \citealt{Drury1983}). 
Relativistic particles in the shocked region emit non-thermal (NT) radiation. In particular, relativistic electrons produce synchrotron emission when they interact with the magnetic fields present in the shocks \citep[e.g.][]{White1985}. The detection of this synchrotron emission in the radio domain constitutes an indirect but unambiguous evidence of particle acceleration. 

The energy budget for particle acceleration depends on the stellar wind kinetic power ($P_{\rm kin}$\footnote{For a star with mass-loss rate of $\dot{M}$ and terminal velocity of $v_\infty$, $P_{\rm kin} = 0.5\,\dot{M}\,v_{\infty}^2$.}). 
Consequently, in recent years, studies have focused on O-type and Wolf-Rayet (WR) systems, which typically have $P_{\rm kin} > 10^{36}$~erg\,s$^{-1}$. 
These investigations detected synchrotron emission in the colliding wind region of binary systems, revealing a strong correlation between particle acceleration and binarity \citep{Dougherty2000, DeBecker2013}. 
To date, approximately 50 NT radio-emitting binary systems have been identified, which are classified as particle-accelerating colliding-wind binaries (PACWBs; \citealt{DeBecker2013,DeBecker2017}). 
On the other hand, for single stars, stellar bow shocks produced by runaway massive stars also present a conducive environment, although clear detections of synchrotron emission have been reported in only three such systems to date \citep{Benaglia2010, Moutzouri2022, van_den_Eijnden2024}. 
For non-runaway stars, so far, NT emission is detected in only one bubble, G2.4+1.4 around WR\,102 \citep{Prajapati2019}. 
This observational breakthrough opens a new window for studying particle acceleration and physical processes associated with shocks in stellar bubbles. 
These observational studies, along with several theoretical works \citep[e.g.,][and reference therein]{Seo2018,Meyer2024,DeBecker2024}, highlighted the ability of massive stars to accelerate particles, positioning them as potential sources of Galactic cosmic rays (GCRs). 
However, the potential of single stars as particle accelerators and their significance as relevant sources of GCRs remains largely unexplored.

\begin{figure*}
    \centering
    \includegraphics[width=\hsize]{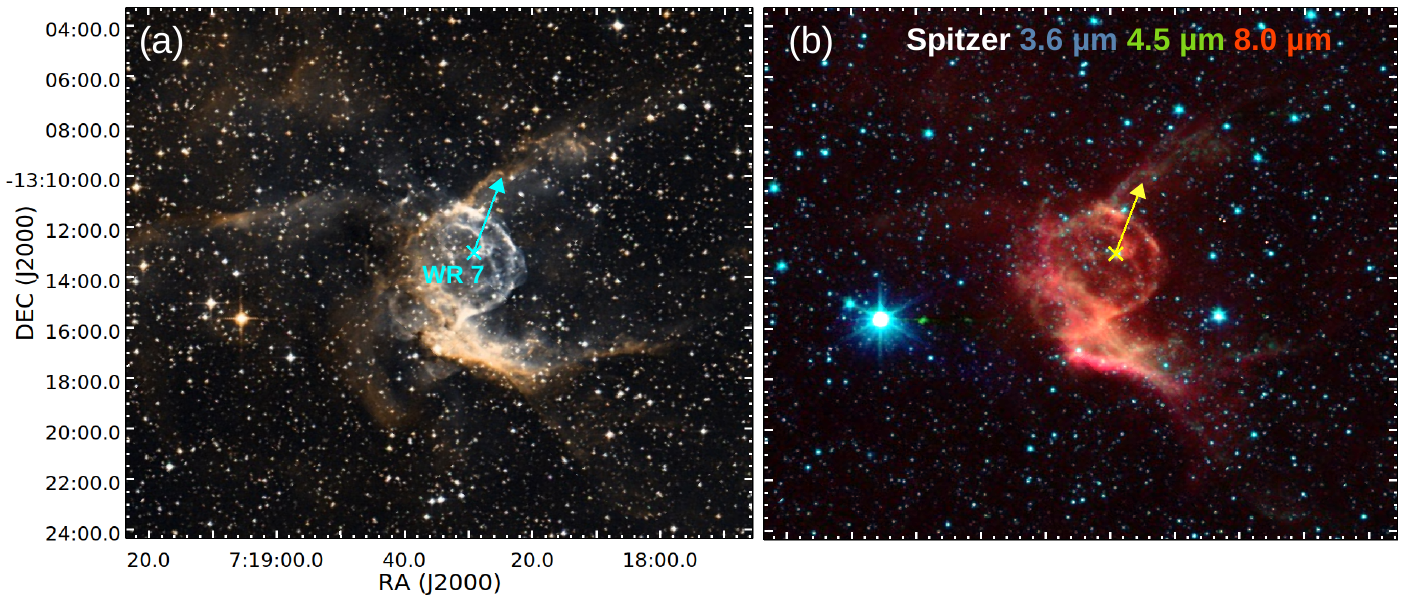}
    \caption{Maps of \ngc in optical and IR. (a) Optical Morphology of \ngc from the Digitized Sky Survey 2 (DSS2). (b) \textit{Spitzer} colour-composite (3.6 $\mu$m (blue), 4.5 $\mu$m (green), and 8.0 $\mu$m (red) bands) image. 
    The arrow indicates the direction of motion of WR\,7 with respect to the local medium (see Section \ref{sec:results-morphology} for details). 
    } 
    \label{fig:WR7-optical-IR}
\end{figure*}

In this context, the bubbles surrounding WR stars are especially important for studying the interaction between the stellar wind and the ambient ISM, given that WR winds typically exhibit $P_{\rm kin} > 10^{37}$ erg s$^{-1}$. 
The termination shocks in these wind-blown bubbles (WBBs) can create the required conditions to drive efficient particle acceleration.  
Studies by \citet{Toala2011,Toala2016} on the evolution of circumstellar material around WR stars provide a detailed overview of WBB formation due to wind--ISM interaction. Recent gamma-ray detections highlight stellar wind termination shocks in single and isolated WR environments as potential particle acceleration sites, where emission likely arises from GCR interactions with nearby matter or radiation fields \citep{Inventar2025}.
In the early evolutionary phase, thermal overpressure and stellar winds drive the expansion of the H{\,\small II} region. 
As the stellar wind propagates through the surrounding ionized gas, it generates shocks that sweep up material, sculpting a bubble. 
By the end of the main sequence (MS) phase, the massive star is typically encompassed by a low-density, hot bubble bounded by a dense shell of neutral material, which expands at a few $\rm km\,s^{-1}$ \citep{Arthur2007}. 
After the MS phase, the mass-loss rate increases, and the star evolves into either a Red Supergiant or Yellow Supergiant or Luminous Blue Variable (LBV) stage, depending on its initial mass \citep{Meynet2003}. 
These stages are characterized by very high mass-loss rates ($10^{-4}$--$10^{-3}$\Msunyr, \citealt{Nugis2000}) and slow winds ($v_{\infty} = 10$--$100$\kms). 
As a result, the star can lose a significant fraction of its mass (reaching up to 50\%) through episodic, non-spherical outbursts of material \citep{Humphreys2010}. 
In the final stage of stellar evolution, as the star enters the WR phase, it generates a powerful wind with terminal velocities between 700 and 5000\kms and a mass-loss rate of $\sim (1$--$5) \times 10^{-5}$\Msunyr \citep{Nugis2000,Sander2012,Sander2019}. 
This fast wind sweeps up and compresses the material expelled during earlier evolutionary stages, thereby forming a WR nebula. 

Here, we aim to investigate the role of the local ISM and its interaction with stellar winds as a necessary ingredient for accelerating particles in single massive stars and to assess their potential as relevant sources of GCRs. For this, we carried out low-frequency (250--500~MHz and 550--950~MHz) radio observations of the WR bubble \ngc encompassing WR\,7. 

This paper is organized as follows. 
Section \ref{sec:ngc2359} gives a brief overview of the WR bubble \ngc.
Section \ref{sec:obs-radiodata} describes the upgraded Giant Meterwave Radio Telescope
(uGMRT) observations, the complementary archival radio datasets used in this study, along with the data reduction process. 
The main results are presented in Section \ref{sec:results}, followed by a detailed discussion of these results and their implications in Section \ref{sec:discussion}. 
Finally, Section \ref{sec:conclusion} summarizes our key conclusions.

\section{WR bubble NGC\,2359}
\label{sec:ngc2359}
With an angular size of 7.4\arcmin, this bubble presents a complex morphology in the IR and optical (shown in Figure~\ref{fig:WR7-optical-IR}), indicative of wind--ISM interaction regions. 
As discussed in \citet{Schneps1981}, \ngc presents a filamentary, nearly spherical shell morphology with multiple streamers and a prominent bright region in its southern part. 
WR\,7 is classified as WN4b type \citep{Smith1996} and exhibits a terminal velocity of 1600\kms with a mass-loss rate of $4 \times 10^{-5}$\Msunyr \citep{Hamann1998}, resulting in a wind kinetic power of $3.3\times 10^{37}\,\rm erg\,s^{-1}$. 
In the literature, different kinematic distances have been adopted for \ngc, ranging from 4.1 to 6.3~kpc. 
Furthermore, the photometric distance to WR\,7 varies between 3.5 and 6.9~kpc \citep[for a more in-depth discussion, we refer to][]{Goudis1994}. 
In addition, the parallax measured by \textit{Gaia} \citep{GaiaDR32022} places the WR star at a distance of 4.8$\pm$0.6~kpc. Considering these estimates, we have adopted a distance of 4.8~kpc for our study. Several parameters of WR\,7 and the bubble are listed in Table \ref{tab:WR7params}.

\begin{table}
    \centering
    \caption{Parameters of WR\,7 and \ngc.}
    \begin{tabular}{lc}
    \hline\hline
    Parameter & Value \\ 
    \hline
    Spectral type\tablefootmark{a} &  WN4b \\ 
    Distance (kpc)\tablefootmark{b} & 4.8$\pm$0.6\\
    $\mu_{\alpha}$ ($\rm mas\,yr^{-1}$)\tablefootmark{b} & $-3.60\pm0.02$\\
    $\mu_{\delta}$ ($\rm mas\,yr^{-1}$)\tablefootmark{b} & $2.99\pm0.02$\\
    $V_{\rm r}$ (\kms) & $-54.4\pm16.8$ \\
    $V_{\rm t}$ (\kms) & $49.8\pm6.6$ \\
    $\theta$ ($^\circ$) & $-19.9^\circ\pm1.3^\circ$ \\
    $\dot{M}$ ($M_{ \odot}\,\mathrm{yr}^{-1}$)\tablefootmark{c} & $4 \times 10^{-5}$ \\
    $v_{\infty}\,\rm(km\,s^{-1})$\tablefootmark{c} & 1600\\
    $P_\mathrm{kin}$ (erg\,$\rm s^{-1}$) & $3.3\times 10^{37}$\\
    Bubble size (\arcmin) & 7.4\\
    \hline
    \end{tabular}
    \label{tab:WR7params}
    \tablefoot{\tablefoottext{a}{\citet{Smith1996}} \tablefoottext{b}{Gaia DR3 data \citep{GaiaDR32022}.} \tablefoottext{c}{\citet{Hamann1998}.} The radial ($V_{\rm r}$) and tangential velocity ($V_{\rm t}$) are corrected for Galactic rotation and $\theta$ indicates the direction of the star's proper motion (measured from north to east).} 
\end{table}

Several studies have examined the distribution and characteristics of the molecular material associated with WR\,7 and \ngc and probed the interaction between ionized and neutral gas using radio and spectral line observations. 
\citet{Cappa1999} reported a strong similarity between the continuum and the line emissions of the nebula using VLA radio continuum (1465~MHz) and HI 21~cm observations of \ngc, with angular resolutions of approximately 30\arcsec and 45\arcsec, respectively. 
Assuming that the radio emission is free--free, the ionized mass of the filamentary shell region was estimated to be 70\,$ M_{\odot}$, and for the entire complex, including the filaments, the estimated mass was in the range of 900--1100~$M_{\odot}$, adopting a distance of 5~kpc. 
This large ionized mass suggests that the shell predominantly consists of interstellar matter. 
The gas kinematics of \ngc have also been studied using CO molecular transitions at different resolutions \citep{Schneps1981,Cappa2001,Rizzo2001,Rizzo2003}.
The CO (1--0) observations from these studies revealed similar results of enhanced molecular line emission at velocities of 37, 54, and 67\kms.  
\citet{Rizzo2001} reported that the surrounding molecular cloud at a velocity of 67\kms was accelerated (from 67 to 53\kms) by at least 14\kms, likely due to interaction with stellar winds. 
In a follow-up work, using higher-resolution CO and $^{13}$CO data, \citet{Rizzo2003} identified three distinct velocity components. 
They analyzed their spatial distribution and properties and suggested that multiple layers of shocked molecular gas in the WBB are associated with episodic interactions of the central star with the circumstellar medium, likely created during the earlier LBV stage and/or the current WR phase of WR\,7.

The complex filamentary bubble morphology and multiple velocity components observed in CO spectra are likely related to episodic mass loss during previous evolutionary phases and shock-driven interactions. 
Additionally, X-ray observations of \ngc reveal high plasma temperatures \citep[$\sim$10$^6$--$10^7$\,K;][]{Toala2015Xray}. This is consistent with the predictions of models for hot shocked stellar wind bubbles, where the stellar wind impacts wind material from a previous evolutionary phase or the ISM and forms an adiabatically shocked region of gas with temperatures ranging from $10^7$~K to $10^8$~K. 
Together with the high wind kinetic power of WR\,7, these factors suggest that shocks in \ngc can provide the necessary conditions for accelerating particles.
%
%
\begin{table}
    \centering
    \caption{Parameters of the radio datasets. }
    \begin{tabular}{ccccc}
    \hline \hline
    & Central Freq & Angular resolution & \textit{rms} & LAS \\    
    & (GHz) & (\arcsec $\times$ \arcsec) & $(\rm \mu Jy\,beam^{-1})$ & (\arcmin)\\ 
    \hline
    \multicolumn{5}{c}{GMRT} \\
    & 0.150\tablefootmark{a} & 29.1 $\times$ 24.9 & 3406 & 68\\
    & 0.402\tablefootmark{b} & 8.3 $\times$ 5.4 & 210    & 25\\  
    & 0.622\tablefootmark{c} & 5.6 $\times$ 4.0  & 17.2 & 16\\ 
    & 0.657\tablefootmark{c} & 5.2 $\times$ 3.8  & 18.4 & 16\\
    & 0.692\tablefootmark{c} & 4.9 $\times$ 3.2  & 16.0 & 15\\
    & 0.728\tablefootmark{c} & 4.8 $\times$ 3.1  & 17.4 & 14\\
    & 0.735\tablefootmark{b} & 5.2 $\times$ 3.7 & 56.0 & 14\\
    & 0.761\tablefootmark{c} & 4.7 $\times$ 3.6  & 20.6 & 13\\
    & 0.801$^c$ & 4.5 $\times$ 3.2  & 25.1 & 13\\ 
    \hline
    \multicolumn{5}{c}{ASKAP} \\ 
    & 0.887 & 13.3 $\times$ 11.2 & 249 & 53\\ 
    & 0.943 & 11.8 $\times$ 10.3 & 191 & 50\\ 
    \hline
    \multicolumn{5}{c}{VLA} \\
    & 1.425 & 21.5 $\times$ 12.9 & 159 & 21\\ 
    & 4.860 & 6.5 $\times$ 3.9   & 99 & 6\\ 
    & 8.689 & 11.9 $\times$ 7.2  & 184 & 3\\ \hline
    \end{tabular}
    \label{tab:WR7-param-radiodataset}
\tablefoot{ \tablefoottext{a}{Map obtained from TGSS survey.} \tablefoottext{b}{ Maps obtained using \texttt{CAPTURE} for Band 3 and Band 4.}  \tablefoottext{c}{Maps of each sub-band of Band 4.}  Global \textit{rms} values are mentioned here. LAS = $\dfrac{\lambda}{B_{\rm min}}$, where $B_{\rm min}$ is the minimum antenna separation. LAS can be significantly lower than these values for data from the snapshot surveys (TGSS and RACS).}
\end{table}

\section{Radio datasets}
\label{sec:obs-radiodata}

We probed the nature of the radio emission and the complex morphologies associated with \ngc using our uGMRT observations at Band 3  (250--500~MHz) and Band 4 (550--950~MHz), along with complementary archival radio datasets at different frequencies.  
The properties (central frequency, resolution, \textit{rms}, and the largest angular structure (LAS)) of the radio maps are compiled in Table \ref{tab:WR7-param-radiodataset}. 
To address the `missing flux' problem in interferometric observations, we estimate the LAS, which is the largest coherent structure that can be imaged by an interferometer and is determined by the length of the shortest baseline. 
In addition, for all radio observations, we adopt a conservative systematic flux uncertainty due to absolute flux calibration of 10\%. 

\subsection{uGMRT observations}
\label{sec:WR7-uGMRT-dataanalysis}

We measured the continuum emission associated with \ngc using the uGMRT located in Pune, India. 
The GMRT employs a hybrid Y-shaped array configuration with 30 antennas of 45-m diameter; 12 antennas are randomly placed within a compact central region spanning 1 km$^{2}$, while the remaining 18 antennas are uniformly distributed along the three arms. 
This hybrid configuration, with baselines ranging from 100\,m to 25\,km, enables the study of ionized emission across both large and small spatial scales.  
An overview of the GMRT system is available in \citet{Swarup1991}, with details of the upgrades presented in \citet{Gupta2017}.
Dedicated observations were conducted in Bands 3 and 4 using the GWB correlator, configured to process 2048 and 4096 channels, respectively.
The primary calibrator (3C147) was observed at the beginning and end of the observation sessions for flux and bandpass calibration. 
The phase calibrator (0744--064) was observed after each 30-minute scan of the target to correct for phase and amplitude variations throughout the observing period.

We processed the full-band data in Bands 3 and 4 using the \texttt{CAPTURE} continuum imaging pipeline for uGMRT \citep{KaleIshwara2021}. 
It uses tasks from Common Astronomy Software Applications (CASA; \citealt{McMullin2007}) for the flagging, calibration, imaging, and self-calibration processes. 
The \citet{PerleyButler2017} scale was used to set the flux density calibration. 
After the initial rounds of editing and calibration, we used the multiterm multifrequency synthesis (MT-MFS; see \citealt{RauCornwell2011}) algorithm in the \textit{tclean} task to account for possible deconvolution errors in wide-band imaging. 
Five rounds of phase-only and one round of amplitude-phase self-calibration were conducted to obtain the final images of the full-band data. 
We note that the residual map for Band 3 (not shown here) indicates that we could not fully recover the flux from the larger scale structures (due to large phase variation and broad radio frequency interference), suggesting that the flux estimate in this band is a lower limit.

To investigate the in-band spectral index in Band 4, we employed the standard procedures of data reduction using the Astronomical Image Processing System (AIPS) for generating sub-band images. A recent paper by \citet{Rashid2024} elaborates on the reliability and accuracy of spectral index estimation using sub-band splitting method. 
The Band 4 $uv$ data were initially divided into six sub-bands, each with a bandwidth of approximately 35~MHz, centered at 622, 657, 692, 728, 761, and 801~MHz. Each sub-band was then carefully examined for corrupted data, including non-functional antennas, bad baselines, and radio frequency interference (RFI). Calibration and imaging were performed separately for each sub-band. 
Channel averaging was optimized to have negligible bandwidth smearing. Further, wide-field imaging techniques were used to account for w-term effects \citep{Cornwell1992}. Several iterations of self-calibration were performed to correct for phase errors and improve the image quality. 
We did not carry out sub-band imaging of Band 3 data 
as it (i) suffered from broad RFI, (ii) showed large phase variations, and (iii) had a relatively smaller bandwidth of $\sim$200~MHz compared to Band~4.
All images used in our analysis were primary beam-corrected.

\subsection{Archival radio data}
To complement our targeted uGMRT observations, we utilize several archival radio data sets at different frequencies. These include radio maps at (i) 150~MHz from the TIFR GMRT Sky Survey (TGSS, \citealt{Intema2017}), (ii) 887 and 943~MHz from the Rapid Australian Square Kilometre Array Pathfinder (ASKAP) Continuum Survey (RACS, \citealt{McConnell2020}), and 
%
(iii) 1425, 4860, and 8689~MHz from the National Radio Astronomy Observatory VLA Archive Survey (NVAS). 
For these datasets, we extracted pipeline-processed images from the respective survey portals\footnote{The TGSS, RACS, and NVAS radio images can be browsed through \url{https://vo.astron.nl/tgssadr/q_fits/imgs/form}, \url{https://data.csiro.au/domain/casdaObservation}, and \url{http://www.vla.nrao.edu/astro/nvas/}, respectively.}. 
The parameters of the archival radio datasets are listed in Table \ref{tab:WR7-param-radiodataset}.

Regarding the RACS data, it is worth noting that we have not included available data at frequencies 1367 and 1655~MHz in our analysis. As discussed in \citet{Duchesne2023,Duchesne2025}, (i) both frequencies are significantly affected by broadband RFI and (ii) there is a decrease in flux density with increasing source size due to incomplete \textit{uv} coverage. The authors mention that baselines below 75~m (for 1367~MHz) and 100~m (for 1655~MHz) were excluded while creating the maps. 
Additionally, \citet{Duchesne2023} highlight the challenges in imaging fields containing sources with extended emission in the Galactic Plane (as is the case with \ngc). These include an increased number of artefacts localised to extended sources, higher \textit{rms}, and a significant portion of the flux density not recovered from them.

\section{Results}
\label{sec:results}
In this study, we primarily focus on determining the nature of the radio emission in \ngc by estimating the spectral index and searching for evidence of NT emission. The spectral index helps us to understand the radiation mechanisms and decipher the underlying physical processes.
\subsection{Morphology of the source}
\label{sec:results-morphology}
\begin{figure*}
    \centering
    \includegraphics[width=\hsize]{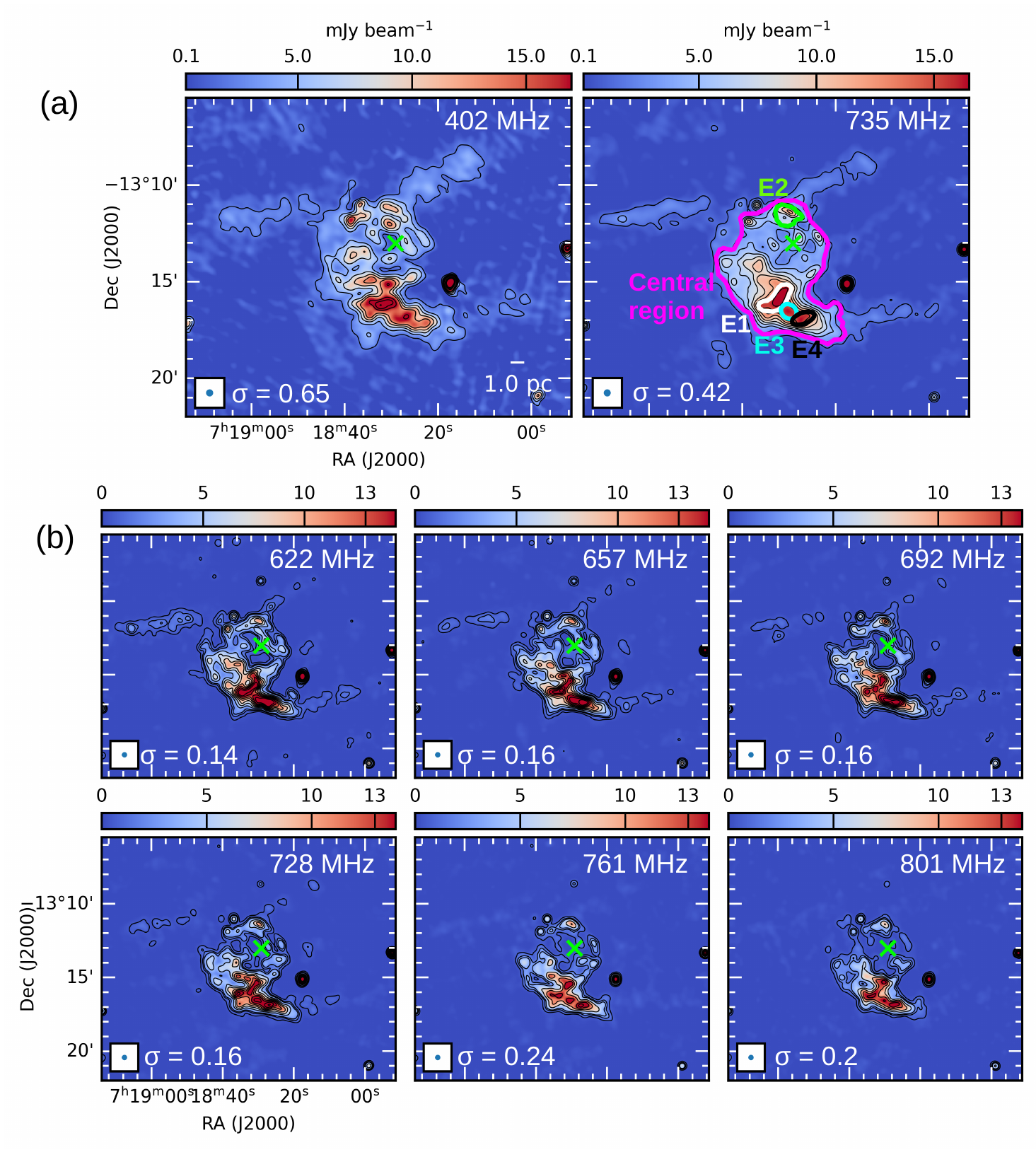}
    \caption{Radio maps of \ngc obtained using our uGMRT data. (a) Maps obtained using the full-band data. (b) Sub-band images for Band-4 data. All maps are convolved to a circular beam of 22\arcsec. For each map, the central frequency and \textit{rms} noise ($\sigma$) in units of $\rm mJy\,beam^{-1}$ are mentioned in each panel. The `X' marks the location of the star in all the panels. Identified apertures (E1--E4) and the central region for obtaining the SEDs are shown in the 735~MHz map.  
    Contour levels are as follows: (i) For maps with central frequencies of 402 and 735~MHz, contour levels start from 3$\sigma$ and increase in steps of 4$\sigma$. (ii) The maps with central frequencies of 622, 657, 692, and 728~MHz, have contour levels of [3, 8, 13, 28, 43, 58, 73, 88, 103, 118] $\times\,\sigma$. (iii) For maps with central frequencies of 761 and 801~MHz, the contour levels are [3, 8, 13, 28, 43, 58, 73] $\times\,\sigma$. In all panels, the contours are smoothed over 3 pixels using a Gaussian kernel.} %
    \label{fig:WR7-ALL-uGMRT-radiomaps}
\end{figure*}

\begin{figure*}
    \centering
    \includegraphics[width=\hsize]{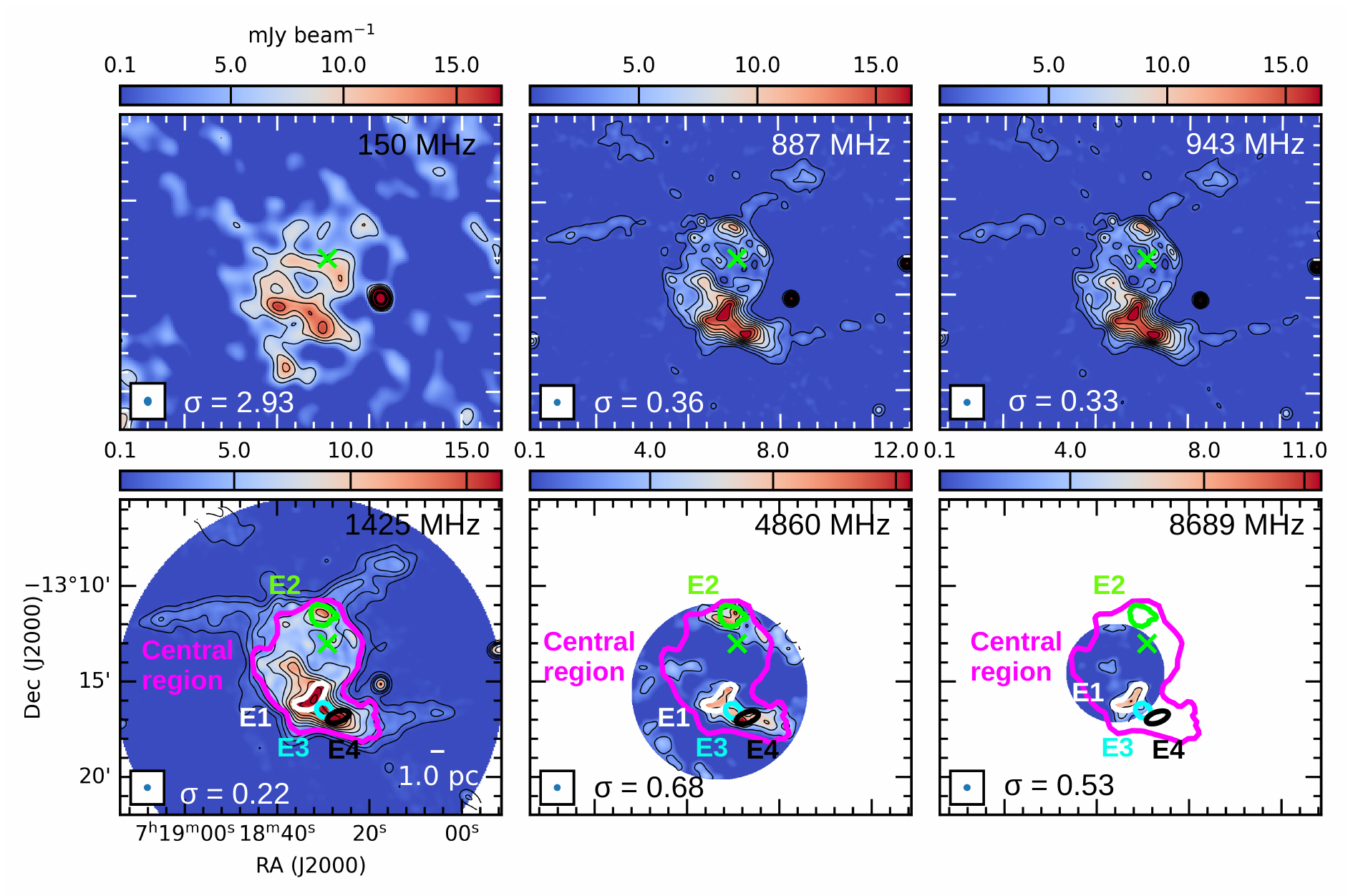}
    \caption{Radio maps of \ngc obtained using archival data. For each map, the central frequency and \textit{rms} noise ($\sigma$) in units of $\rm mJy\,beam^{-1}$ are mentioned in each panel. Barring the image at 150~MHz, all maps are convolved to a circular beam of 22\arcsec. The 150~MHz map has a beam size of 29.1\arcsec $\times$ 24.9\arcsec and is smoothed across 3 pixels using a Gaussian kernel. The `X' marks the location of the star in all the panels. Identified apertures (E1--E4) and the central region for obtaining the SEDs are shown in the 1425, 4860, and 8689~MHz maps. 
    Contour levels are as follows: (i) For maps with the central frequency of 150~MHz, contour levels start from 2$\sigma$ and increase in steps of 1$\sigma$. (ii) For maps with central frequencies of 887 and 943~ MHz, contour levels start from 3$\sigma$ and increase in steps of 5$\sigma$, (iii) The 1425~MHz map has contour levels of [3, 7, 11, 15, 23, 31, 39, 47, 54, 62, 70, 78] $\times\,\sigma$. (iv) For maps with central frequencies of 4860 and 8689~MHz, contour levels start from 3$\sigma$ and increase in steps of 4$\sigma$. In all panels, the contours are smoothed over 3 pixels using a Gaussian kernel.}
    \label{fig:WR7-ALL-archive-radiomaps}
\end{figure*}

Radio continuum maps from GMRT, ASKAP and VLA are shown in Figures~\ref{fig:WR7-ALL-uGMRT-radiomaps} and \ref{fig:WR7-ALL-archive-radiomaps}. 
All maps are convolved with a common circular beam size of 22\arcsec  to compare the emission at different frequencies, except for the map at 150~MHz that has a poorer resolution. Almost all radio maps trace a similar complex morphology as seen in optical and IR (Figure~\ref{fig:WR7-optical-IR}). 
The southern part of the complex shows strong radio emission. In the maps from the six sub-bands, the diffuse emission near WR\,7 is observed to be weak, especially at 761 and 801~MHz, compared to the full-band 735~MHz image, even though the resolution and \textit{uv} coverage are similar. This is possibly because the sub-band images have much smaller bandwidths than the full-band \texttt{CAPTURE} maps, which reduces their sensitivity to fainter, extended emission. The radio morphology reported in \citet{Cappa1999}, where this source was mapped using VLA data at 1465~MHz 
with a lower resolution of $\sim$ 39\arcsec $\times$ 25\arcsec, is similar to the uGMRT maps. Due to lower resolution, more extended diffuse emission is traced in their map (see Figure 2 of \citealt{Cappa1999}). 
The maps reveal an asymmetrical bubble structure around WR\,7, which is offset from the centre. A similar picture is also seen in the bubble, G2.4+1.4, where the star WR\,102 is observed to be offset from the bubble centre \citep{Prajapati2019}. In a previous study of G2.4+1.4, \citet{Dopita1990} attributed this asymmetry to Rayleigh-Taylor instabilities, proposing that the location of WR\,102 near a molecular cloud restricts expansion to that side while allowing expansion on the other side into a low-density ISM, resulting in a ``scalloping''  morphology. 
For \ngc, CO observations \citep{Schneps1981,Rizzo2001,Rizzo2003} have revealed bright extended emission in the southern part, supporting the above explanation.

\begin{figure}
    \centering
    \includegraphics[width=\hsize]{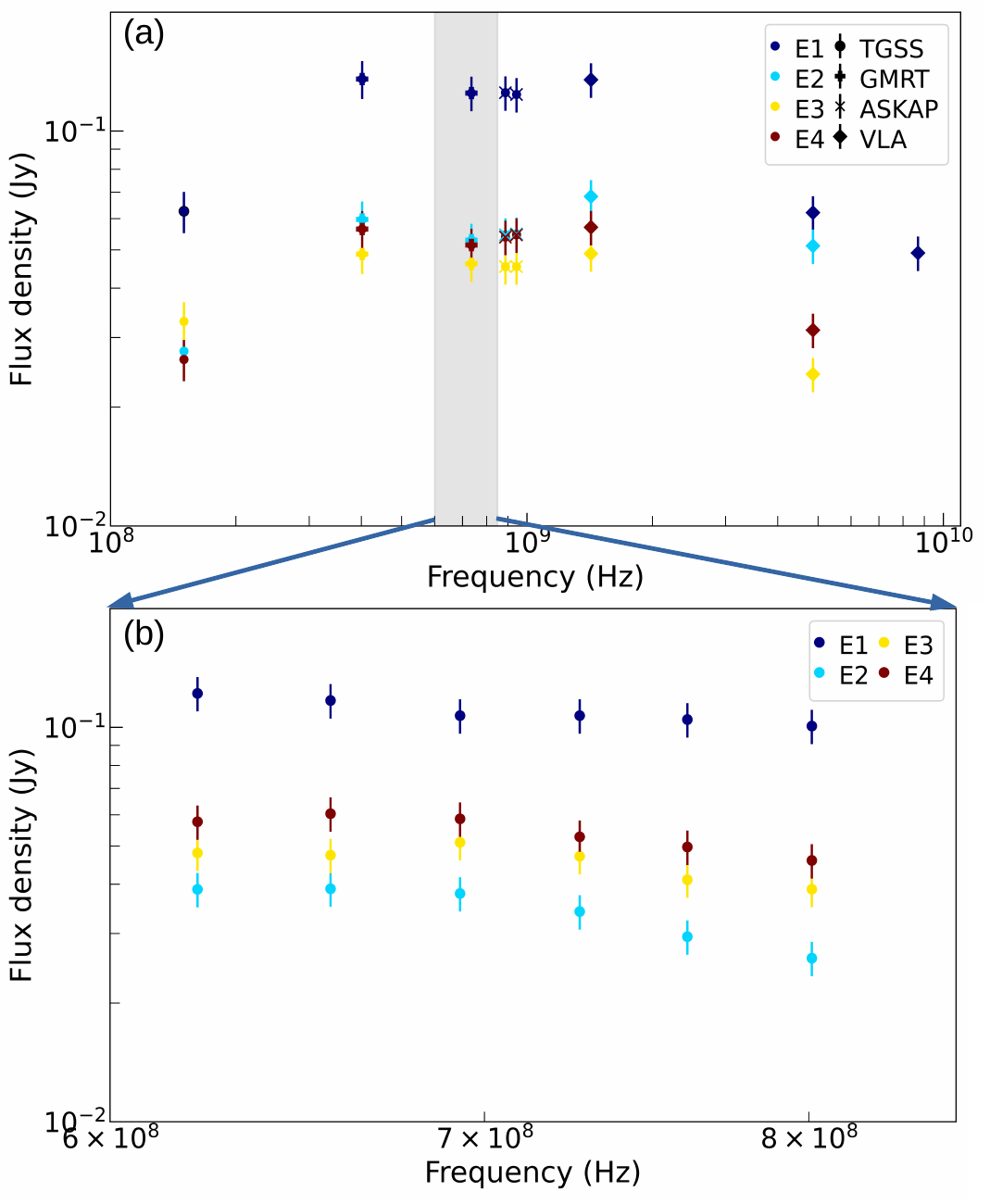}
    \caption{Radio SED for the identified apertures (E1--E4) shown in Figures \ref{fig:WR7-ALL-uGMRT-radiomaps} and \ref{fig:WR7-ALL-archive-radiomaps}. 
    (a) SED using the maps at 150, 402, 735, 887, 943, 1425, 4860, and 8689~MHz maps.
    The data point at 402~MHz (Band 3) is a lower limit (see Section~\ref{sec:WR7-uGMRT-dataanalysis} for details). (b) SED using the maps of the six sub-bands (622, 657, 692, 728, 761, and 801~MHz) of the Band 4 GWB uGMRT data. The uncertainty in the flux density measurements is estimated using \citet{Lal2021} $\Delta S = [ (S_\nu \times f )^{2} + rms^{2} \times N_{\rm beams} ]^{0.5}$, where $S_\nu$ is the flux density, $f$ is an absolute flux density calibration uncertainty (taken as 10\%), and $N_{\rm beams}$ is the number of synthesized beams in the aperture.}
    \label{fig:WR7-SED}
\end{figure}

Additionally, several simulation studies \citep[e.g.,][]{Brighenti1995a,Brighenti1995b,Iturbide2019} have shown that a high proper motion of the star leads to an asymmetry in the bubble structure. 
To investigate this possibility, we estimated the stellar velocity with respect to its surrounding medium by correcting the proper motions measured by \textit{Gaia} for Galactic rotation as in \cite{Martinez2023}\footnote{\url{https://github.com/santimda/intrinsic_proper_motion}}. The radial ($V_{\rm r}$) and tangential ($V_{\rm t}$) velocities, and the direction of motion ($\theta$), are provided in Table \ref{tab:WR7params}. This implies that WR\,7 is moving with a supersonic velocity in the direction shown in Figure~\ref{fig:WR7-optical-IR}, which is the projection of its motion onto the plane of the sky and explains the off-centered location of WR\,7. 
Moreover, such a high peculiar velocity should shape the WBB into a more paraboloidal structure known as a bow shock \citep{Weaver1977}. 
The apex of the bow shock is located in the direction of motion of the star, at a stagnation radius given by the balance of the momentum flux of the wind with the ram pressure of the surrounding ISM.  
The fact that the radio maps do not show any discernible bow shock morphology could be due to the radial velocity being higher than the tangential velocity, though the uncertainty in the velocity estimates is large. 
This results in a projected structure that is more circular, which would be challenging to identify from our radio maps. Furthermore, the medium through which the WR wind propagates might deviate significantly from the standard homogeneous medium assumed in bow shock models. 
Moreover, no bow shocks have been identified around any WR stars, although some of them are classified as runaway stars \citep{Meyer2020}. 
This non-detection of bow shocks is consistent with numerical simulations by \citet{Meyer2020} that study the circumstellar material surrounding a 60 $M_{\odot}$ runaway WR star. The simulations suggest that the interaction of the high-velocity WR phase wind with the slower wind from earlier evolutionary stages leads to the formation of spherical shells. These shells seem to comove with the star regardless of its proper motion or the properties of the surrounding ISM, and consequently, no bow-shaped region is identified. Furthermore, \citet{Meyer2020} proposed that a low-density medium around these stars can explain the absence of bow shocks in runaway stars like WR\,71, WR\,124, and WR\,148. 
For \ngc, without knowledge of the density, it is difficult to draw a conclusion on the reason for the non-detection of a bow shock. Additional studies to estimate the density of the ISM are required for further investigation. 

\subsection{Nature of radio emission: Spectral index}
\label{sec:results-spectralindex}
The spectral index ($\alpha$) is defined as $S_{\nu} \propto \nu^{\alpha}$, where $S_{\nu}$ is the flux density at frequency $\nu$. 
For thermal free--free emission from ionized gas, $\alpha$ ranges from +2 (optically thick) to $-0.1$ (optically thin). The free--free opacity depends on the frequency, which causes sources to shift from being optically thick at lower frequencies to optically thin at higher frequencies. This transition point (turnover frequency) depends on the plasma density. Moreover, the slope of the spectrum can vary depending on the density distribution of the plasma. For instance, in the case of stellar winds of massive stars (with a density distribution that scales with the inverse of the distance to the star), an intermediate value of $\alpha \sim 0.6$ is expected \citep{Wright1975}. 
On the other hand, optically thin synchrotron radiation generally has a steep spectral index of $\alpha \approx -0.5$, assuming high Mach number adiabatic shocks and monoatomic gas. 
However, absorption or suppression processes can modify the observed spectrum, making the spectral index less negative or even positive \citep[e.g.,][]{Melrose1980}.   
A spectral index of $\alpha < -0.1$ strongly indicates NT emission, while positive values typically point to a thermal origin. 
In many astrophysical environments, thermal plasma can coexist with a population of relativistic electrons, leading to spectra that include both thermal and NT components. 
Despite this complexity, it is possible to differentiate between these components by analyzing the spectral energy distribution (SED) across a wide range of frequencies. 
NT emission, with a negative spectral index, usually dominates at lower frequencies, while thermal radiation contributes mainly at higher frequencies \citep[e.g.,][]{DeBecker2018}. 
Therefore, a comprehensive sampling of the SED at low frequencies is necessary to investigate the interplay between absorption and emission processes. 

To estimate the global spectral index, we calculated the integrated flux density within the 5$\sigma$ contour of the 735 MHz map. This aperture encompasses the central region of the bubble, as shown in Figures \ref{fig:WR7-ALL-uGMRT-radiomaps} and \ref{fig:WR7-ALL-archive-radiomaps}, and matches the extent of the field of view (FOV) of the 4860 MHz map. However, as the FOV at 8689 MHz is much smaller than this region, flux was not obtained at that frequency. 
The resultant SED  shows an interesting trend, where the slope is seen to vary: (i) between 150~MHz and 402~MHz, the slope is positive ($\alpha= 0.27\pm0.15$); (ii) between 402~MHz and 1425~MHz, the slope becomes approximately flat within the error bars ($\alpha=-0.01\pm0.17$); and (iii) from $\sim$1.4~GHz to higher frequencies, the slope turns negative ($\alpha=-0.93\pm0.12$). 
To examine the variation of spectral index across the bubble, we identified apertures with bright emissions at all frequencies from 150~MHz to 8.7~GHz, labelled as E1--E4 (see Figures \ref{fig:WR7-ALL-uGMRT-radiomaps} and \ref{fig:WR7-ALL-archive-radiomaps}), and estimated the spectral index from the integrated flux density within each aperture. The SEDs shown in Figure~\ref{fig:WR7-SED} follow the same trend as the global SED of the central region: (i) between 150~MHz and 402~MHz the spectral index is $\alpha= 0.4\,\rm{to}\,0.8$; (ii) between 402~MHz and 1425~MHz, the slope is almost flat; 
and (iii) from $\sim$1.4 GHz to higher frequencies, the slope ranges from $\alpha=-0.2\,\rm{to}\,-0.6$. 
The spectral indices for the above frequency ranges are given in Table~\ref{tab:WR7-spectralindex-sed}.  
Consistent with this, the radio SED presented in Figure~\ref{fig:WR7-SED}(b), derived from the Band 4 uGMRT data for the six sub-bands (622 to 801~MHz) shows a nearly flat trend. As discussed previously, the sub-band images do not fully recover the diffuse emission, and the average flux densities of the apertures for the sub-band images are slightly lower than the values at 735~MHz, especially for E2. This can be attributed to the smaller bandwidth of the sub-band maps, as discussed previously in Section~\ref{sec:results-morphology}. Hence, in our spectral analysis, we use the flux densities derived from the full-band CAPTURE maps and archival radio maps.

\begin{table}[ht]
    \caption{Spectral index values for each segment of the SEDs plotted in Figure \ref{fig:WR7-SED}(a).}
    \centering
    \begin{tabular}{cccc}
    \hline \hline
     Aperture & $\alpha\tablefootmark{a}_{(150/402)}$ & $\langle \alpha_{402-1425} \rangle\tablefootmark{b}$ & $\alpha\tablefootmark{a}_{(1425/4860)}$ \\ 
    \hline
    Central region     & $0.27\pm0.15$   & $-0.01\pm0.17$     & $-0.93\pm0.12$ \\ 
     E1     & $0.78\pm0.14$   & $-0.02\pm0.05$     & $-0.57\pm0.05\tablefootmark{c}$    \\%
     E2     & $0.78\pm0.14$   & $0.08\pm0.12$      & $-0.23\pm0.11$         \\%
     E3     & $0.40\pm0.14$   & $-0.01\pm0.05$     & $-0.57\pm0.11$         \\%
     E4     & $0.77\pm0.14$   & $0.0\pm0.05$       & $-0.49\pm0.11$         \\ 
    \hline
    \end{tabular}
    \label{tab:WR7-spectralindex-sed}
\tablefoot{ \tablefoottext{a}{Spectral indices between the indicated frequencies (in MHz).} \tablefoottext{b}{Average spectral index obtained using values at 402, 735, 887, 943, and 1425~MHz.} \tablefoottext{c}{Spectral index obtained by fitting flux densities at 1425, 4860, and 8689~MHz.} When more than two data points are used for fitting, the error in $\alpha$ reflects the fitting error, while for cases with only two points, the analytical error in $\alpha$ is obtained using the \textit{uncertainties} package in Python.} 
\end{table}

The accuracy of the spectral index estimates strongly depends on the spatial scales probed in the frequency bands.  
Given the large spatial extent of the bubble, we explored whether the flux densities at higher frequencies are affected by missing flux because of filtering out of structures larger than the LAS. 
For this, we compared the FOV, aperture sizes and LAS of the observations at 4860 and 8689~MHz. 
For VLA 4860~MHz (observed in VLA-C configuration), the native resolution, FOV, and LAS are 6.5\arcsec $\times$ 3.9\arcsec, 9.2\arcmin, and 6\arcmin, respectively. For VLA 8689~MHz (observed in VLA-D configuration), they are 11.9\arcsec $\times$ 7.6\arcsec, 5.2\arcmin, and 3\arcmin, respectively.  
On the other hand, the diameter of the entire bubble is 7.4\arcmin, while the central region and the largest aperture (E1), selected for deriving the SEDs have sizes of 5.8\arcmin and 1.3\arcmin, respectively. 
In the VLA 4860~MHz map, the size of the entire bubble is slightly larger than the LAS; the size of the central region is comparable to the LAS, and aperture E1 is around five times smaller than the LAS. Given the small FOV of the VLA 8689~MHz map, we could only obtain the flux for aperture E1, which has a size much smaller than the LAS. Hence, it is unlikely that there is any significant loss of flux in these higher frequencies for the selected apertures.

\section{Discussion}
\label{sec:discussion}
The observed radio SEDs of the bubble, as well as regions within, present distinct slopes in different frequency domains. Evidence of NT emission is seen above 1~GHz, where the spectral index values are negative, even steeper than the canonical value of $-0.5$ for optically thin synchrotron radiation from relativistic electrons accelerated by DSA in high Mach number adiabatic shocks. As we move below 1~GHz to $\sim$400~MHz, the flattening is indicative of a turnover due to absorption/suppression processes at work \citep[e.g.][]{Melrose1980}. To elucidate the turnover mechanism and derive physical information about the system, we need to introduce a radiative model.

\subsection{SED Modelling}

We introduce an SED model that is physically motivated. The emission region is most likely composed of a combination of ionized gas and relativistic particles. Thus, we include both free--free and synchrotron emission components, although the negative spectral indices observed already point to synchrotron being the dominant component. The observed low-frequency synchrotron SED can be affected by: (1) synchrotron self-absorption (SSA), (2) Razin-Tsytovich (RT) effect, and (3) free--free absorption (FFA) by thermal plasma mixed with the relativistic particles (hereafter, internal FFA).
Each of these processes leads to specific spectral features. Under the assumption of a homogeneous region, we discuss the emission and absorption processes in more detail below.
\begin{table*}[h!]
    \centering
    \caption{List of parameters in the model used to fit the SEDs.}
    \begin{tabular}{lccc}
    \hline
    Parameter  & Symbol & \multicolumn{2}{c}{Range}  \\ 
    \cline{3-4} 
              &  & E1--E4 & Central region  \\
    \hline
    Reference frequency &  $\nu_0$ (GHz) & \multicolumn{2}{c}{1}  \\
    Normalization of the synchrotron component &  $A_\mathrm{sy}$ (mJy) & $[10^{-1}, 10^{3}]$ & $[10^{2}, 10^{5}]$ \\
    Spectral index for optically thin synchrotron emission&  $\alpha_\mathrm{sy}$ & $[-1.0, -0.5]$\tablefootmark{a} & $[-1.0, -0.5]$ \\
    Electron number density & $n_{\rm e}$ ($\rm cm^{-3}$) & $[10^{1.0}, 10^{3.0}]$ & $[10^{0}, 10^{2.0}]$  \\
    Magnetic field strength & $B$ ($\mu\rm G$) & $[10^{-0.3}, 10^{1.0}]$ & $[10^{-0.7}, 10^{1.0}]$  \\
    Electron temperature & $T_{\rm e}$ (K) & \multicolumn{2}{c}{$9000$}  \\
    \hline
    \end{tabular}
    \label{tab:model-params}
    \tablefoot{For parameters spanning over a large range, we fit the logarithm of the quantity. \tablefoottext{a}{For aperture E2, E3 and E4, the spectral index is fixed at $-$0.5.}}
\end{table*}
\begin{table*}[h!]
    \centering
    \renewcommand{\arraystretch}{1.5}
    \caption{Best fit parameters from SED modelling.}
    \begin{tabular}{ccccccccc}
    \hline
    \multirow{2}{*}{Apertures}  & \multirow{2}{*}{$\chi^{2}$} & \multirow{2}{*}{d.o.f} &\multicolumn{4}{c}{Best fit values} &\multicolumn{2}{c}{Inferred values}  \\ 
    \cmidrule(lr){4-7} 
    \cmidrule(lr){8-9}
    &  & & $\log A_{\rm sy}$ (mJy) & $\alpha_{\rm sy}$ & $\log{n_{\rm e} \, (\rm cm^{-3})}$ & $\log{B \, (\mu {\rm G})}$ & $\nu_{\rm Razin} \, ({\rm GHz}) $ & $\nu_{\rm FFA}\,({\rm MHz}) $\\
    \hline
     Central region & 13.3 & 3 & $3.11^{+0.04}_{-0.06}$ & $-0.68^{+0.1}_{-0.14}$ & $0.78^{+0.25}_{-0.29}$ & $-0.3^{+0.19}_{-0.27}$ & $0.24^{+0.06}_{-0.05}$ & $12.7^{+9.8}_{-5.9}$\\  
     \hline
     E1 & 5.0 & 4 & $2.21^{+0.04}_{-0.05}$ & $-0.69^{+0.1}_{-0.18}$ & $1.45^{+0.16}_{-0.27}$ & $0.13^{+0.1}_{-0.18}$ & $0.39^{+0.13}_{-0.07}$ & $27.9^{+11.5}_{-12.6}$\\ 
     E2 & 8.1 & 4 & $1.8^{+0.09}_{-0.12}$ & $-0.5\tablefootmark{a}$ & $1.63^{+0.06}_{-0.09}$ & $0.12^{+0.11}_{-0.11}$ & $0.61^{+0.22}_{-0.17}$ & $38.7^{+5.1}_{-6.9}$\\ 
     E3 & 3.5 & 4 & $1.72^{+0.04}_{-0.08}$ & $-0.5\tablefootmark{a}$ & 1.66\tablefootmark{b} & 0.58\tablefootmark{b} & $0.24^{+0.04}_{-0.03}$ & $19.6^{+13.9}_{-8.8}$\\  
     E4 & 3.0 & 4 & $1.83^{+0.04}_{-0.05}$ & $-0.5\tablefootmark{a}$ & 1.62\tablefootmark{b} & 0.4\tablefootmark{b} & $0.31^{+0.05}_{-0.03}$ & $19.2^{+12.6}_{-8.2}$\\         
    \hline
    \end{tabular}
    \label{tab:bestfit-params}
    \tablefoot{Errors correspond to 1$\sigma$ uncertainties based on the 16th and 84th percentiles of the posterior distributions. \tablefoottext{a}{For aperture E2, E3 and E4, we fixed the spectral index at $-$0.5.  
    \tablefoottext{b}{For E3 and E4, $n_{\rm e}$ and $B$ are highly correlated (see Fig.~\ref{fig:E1-4cornerplot}), indicating degeneracy between them. Additionally, the lower limits are unconstrained and can be any arbitrary value; hence, we report their maximum values.}
    }
    } 
\end{table*}
\begin{figure*}
    \centering
    \includegraphics[width=0.48\hsize]{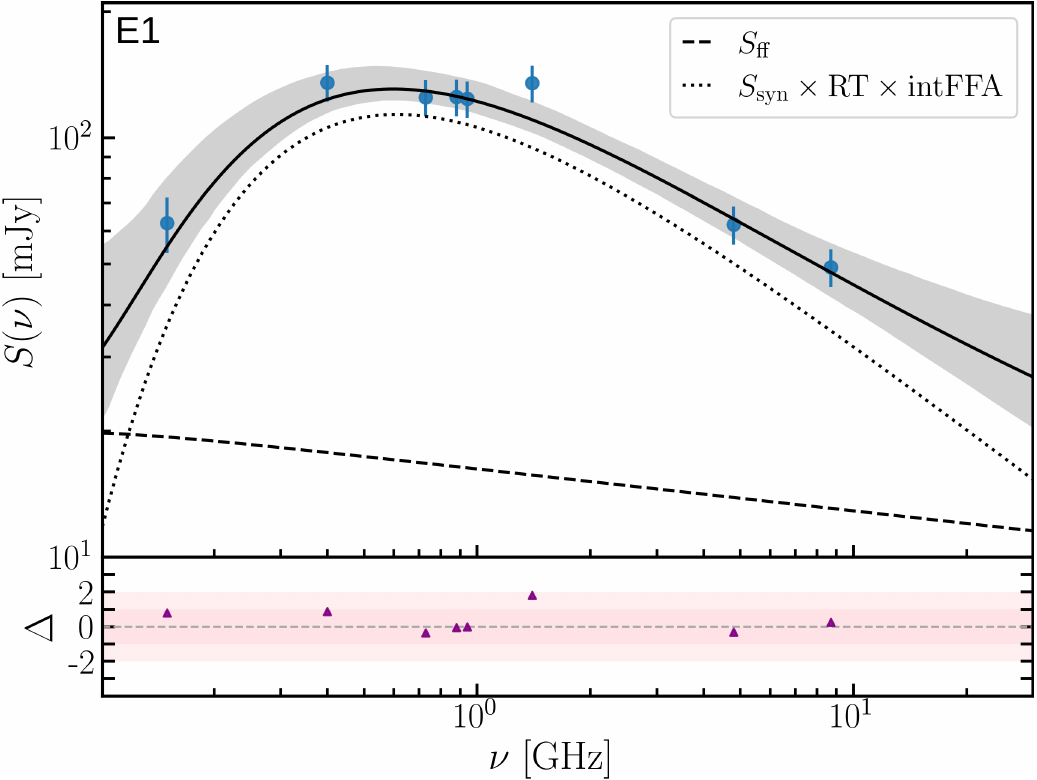}
    \includegraphics[width=0.48\hsize]{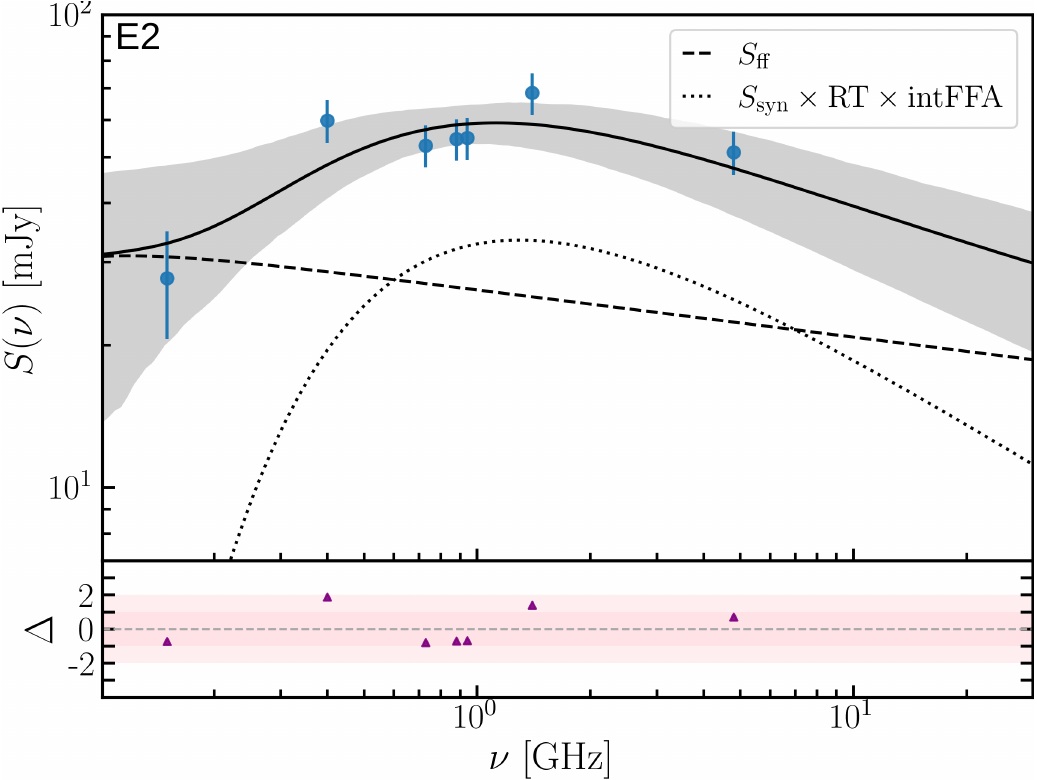}
    \includegraphics[width=0.48\hsize]{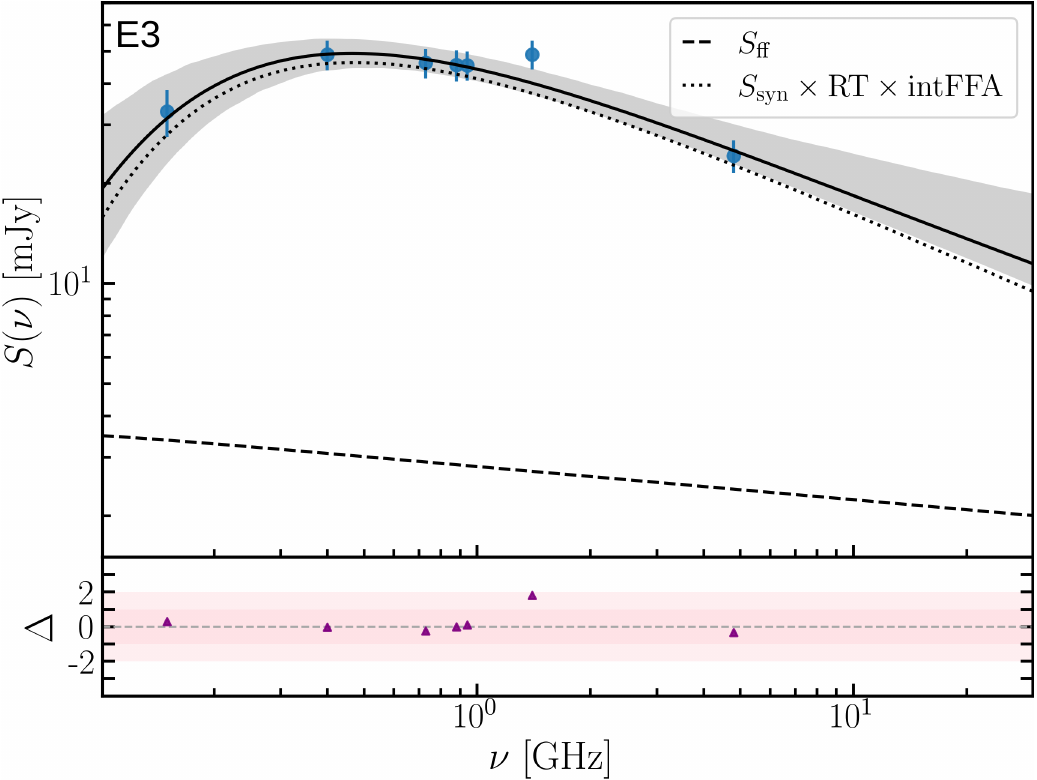}
    \includegraphics[width=0.48\hsize]{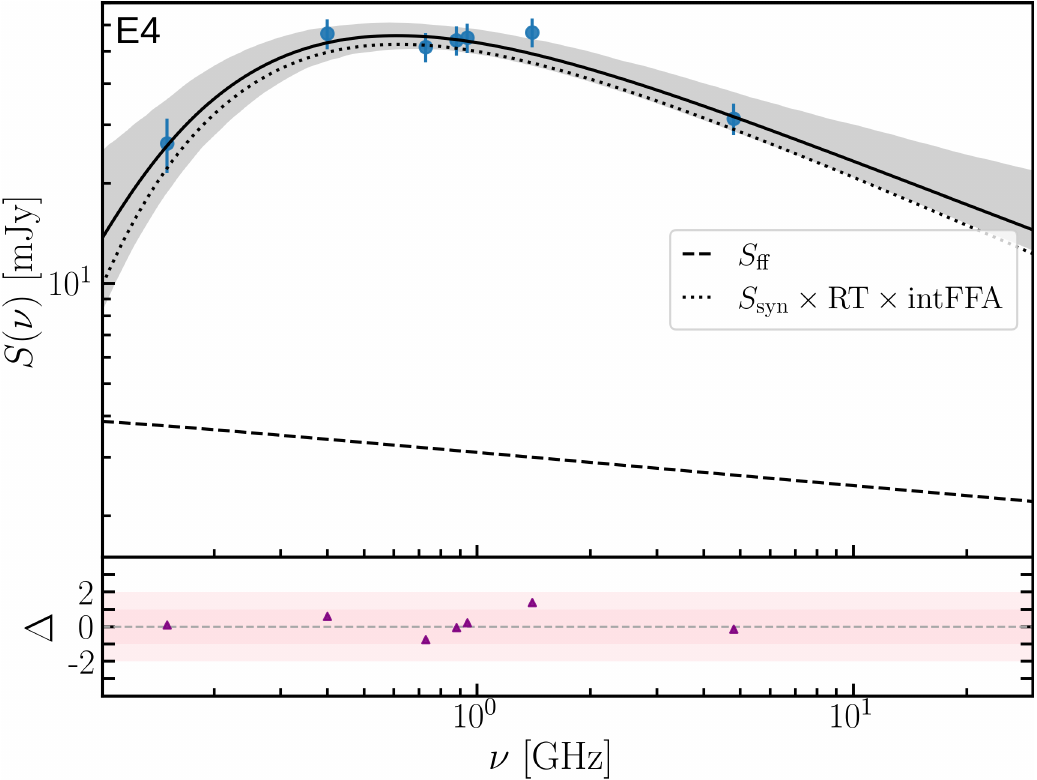}
    \includegraphics[width=0.48\hsize]{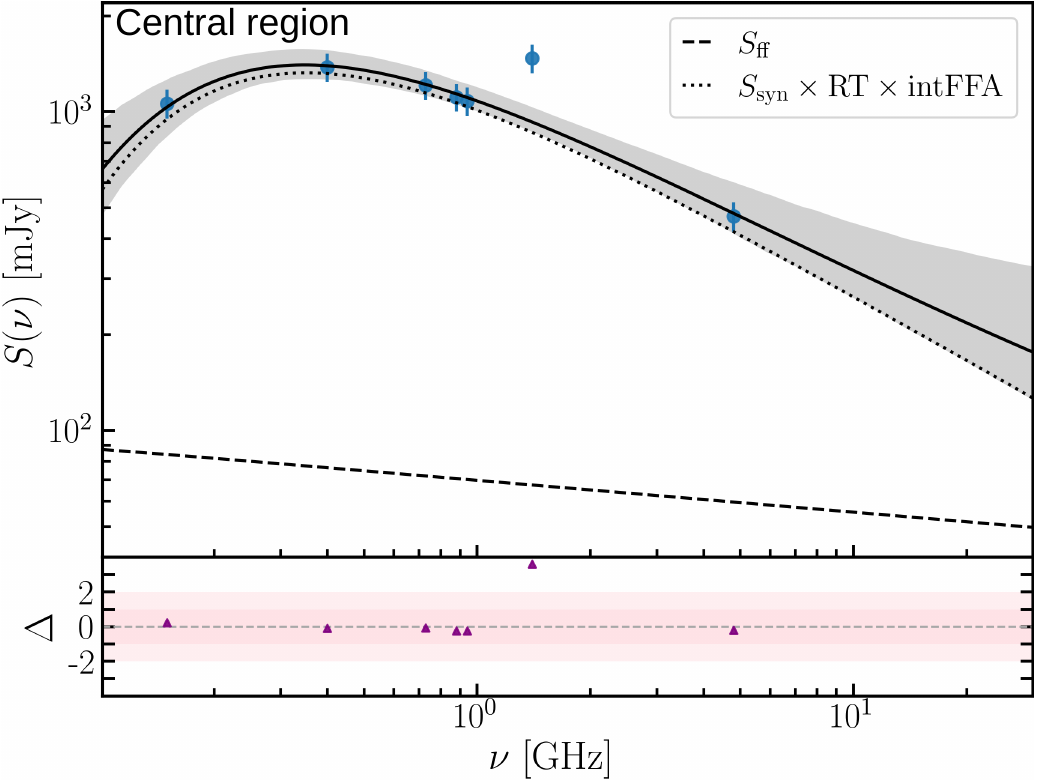}
    \caption{SED fitting for apertures E1--E4 and the central region using the composite model described in Equation \ref{eq:model-composite}. The shaded regions show the 1$\sigma$ confidence interval. The bottom panels of each image present the residuals of the fit. The corresponding corner plots showing the posterior distribution of each parameter are presented in the Appendix \ref{sec:appendix-A}.}
    \label{fig:SED-fits}
\end{figure*}
\subsubsection{Emission processes}
    \begin{itemize}
        \item Synchrotron radiation: The optically thin synchrotron emission arising from a relativistic electron population in the presence of a magnetic field can be written as 
        \begin{equation}\label{Eq:synch}
        S_{\rm{sy}} = A_{\rm sy}\,\left(\dfrac{\nu}{\nu_{0}}\right)^{\alpha_{\rm{sy}}},
        \end{equation}
        where $S_{\rm{sy}}$ is the synchrotron flux density, $A_{\rm sy}$ is a normalization factor, $\alpha_{\rm{sy}}$ is the synchrotron spectral index and $\nu_0$ is a reference frequency. 
        This parameterization encodes information on the relativistic particle population and the magnetic field intensity. Namely, for a power-law electron energy distribution, $N(E) = N_0 \, E^{-p}$, $A_{\rm sy}$ is a function of $N_0$, $p$, and $B$, and $\alpha_\mathrm{sy}=-(p-1)/2$ \citep[e.g.][]{Blumenthal1970}. We can re-write these quantities in terms of energy density in non-thermal electrons, $U_\mathrm{e}$, and in magnetic fields, $U_B$, as
        \begin{equation}
            U_\mathrm{e} \approx \frac{E_\mathrm{min}^{2-p}}{p-2} \frac{N_0}{V}, \qquad U_B = \frac{B^2}{8\pi},
            \label{eq:energy_density}
        \end{equation}         
        where $V$ 
        is the volume of the emitting region and $E_\mathrm{min} \sim 1$~MeV is the minimum energy of the relativistic electrons.

    %
        \item Thermal free--free emission: 
        The thermal free--free flux density can be expressed as \citep{Mezger1967}:
        \begin{equation}\label{Eq:ff}
        S_{\rm ff} = 3.07 \times 10^4\,T_e\,\nu^2\,\Omega\,\left(1 - e^{-\tau(\nu)} \right) ,
        \end{equation}
        where the optical depth ($\tau$) is given by
        \begin{equation}\label{Eq:tau}
        \tau(\nu) = 8.21\, a\,\times\,10^{-2}\,\nu^{-2.1}\,EM\,T_{\rm e}^{-1.35},
        \end{equation}
        and the emission measure ($EM$), and solid angle ($\Omega$) are:
        \begin{equation}
        EM = 2\, n^2_{\rm e} \, R, \qquad
        \Omega = \pi \left( \dfrac{R}{d} \right)^2.
        \end{equation}
        In the above equations, $S_{\rm ff}$ is the integrated flux density in Jy; $T_{\rm e}$ is the electron temperature in K; $\nu$ is in GHz; $EM$ is in cm$^{-6}$\,pc; $n_{\rm e}$ is the electron number density in cm$^{-3}$; 
        $\Omega$ is the solid angle subtended by the source in steradian; $R$ is the radius of the apertures in pc, and $d$ is the distance to the bubble in pc. Also, based on Table~6 of \citet{Mezger1967}, we used a value of 0.99 for the correction factor $a$. 
    \end{itemize}

\subsubsection{Turnover processes}
\begin{itemize}
        \item SSA: this is the process by which synchrotron photons are re-absorbed by the same relativistic electron population that emits them. It leads to an optically thick spectrum with a positive slope ($\alpha=2.5$) below a specific turnover frequency, causing a notable reduction in synchrotron flux density at lower frequencies. 
        To note, SSA is only effective in very compact and luminous radio sources in which the number density of relativistic electrons is high. In contrast, \ngc is an extended structure that spans several parsecs and exhibits relatively low luminosities. 
        Hence, SSA is unlikely to be a significant absorption mechanism in this particular environment and is not considered in our model.
        \item RT effect: The synchrotron radiation produced by relativistic electrons embedded in a thermal plasma is suppressed when the refractive index of the medium is less than unity \citep{Melrose1972}. This is known as the Razin-Tsytovitch effect, 
        and it depends on the plasma frequency, which in turn depends on $n_\mathrm{e}$ and $B$. Observationally, it causes a turnover below the Razin frequency ($\nu_{\rm{Razin}}$) given by \citep{Pacholczyk1970}
        \begin{equation}\label{Eq:razin-fq}
        \nu_{\rm{Razin}} = 20\,\dfrac{ n_\mathrm{e}  }{ B  }\,(\rm MHz) ,
        \end{equation}
        \noindent where $n_{\rm e}$ is the electron number density in $\rm cm^{-3}$ and $B$ is the magnetic field strength in $\mu$G. 
        The effect results in an exponential cut-off in the synchrotron radio spectrum at lower frequencies \citep[e.g.][]{Bloot2022}:
        \begin{equation}\label{Eq:synchRZ}
         S_{\rm{sy}} = A_{\rm sy}\,\left(\dfrac{\nu}{\nu_{0}}\right)^{\alpha_{\rm{sy}}} \, e^{ - \dfrac{\nu_{\rm Razin}}{\nu} } ,
        \end{equation}
        where the parameters have the same meaning as in Eq.~\ref{Eq:synch}. 
        \item Internal FFA: where a thermal plasma and relativistic electrons coexist, the ionized gas can significantly absorb the synchrotron photons as they propagate within the emitting region, causing a turnover in the synchrotron spectrum referred to as internal FFA. 
        In this case, the drop in radio flux can be expressed as \citep{Tingay2003}
        \begin{equation}\label{Eq:synchIFFA}
         S_{\rm{sy}} = A_{\rm sy}\,\left(\dfrac{\nu}{\nu_{0}}\right)^{\alpha_{\rm{sy}}} \,  \frac{1 - e^{-\tau (\nu)} }{\tau(\nu)},
        \end{equation}
        where the parameters have the same meaning as in Eqs.~\ref{Eq:synch} and \ref{Eq:tau}. 
        The occurrence of internal FFA in astrophysical synchrotron sources has been addressed by \citet{deBruyn1976}. 
        Although that study was motivated by the need to provide an adequate interpretation framework for extragalactic sources, the basic principle deserves to be considered for other synchrotron sources. 
        Additionally, internal FFA has more recently been proposed to be responsible for the morphology of the radio SED of the PACWB WR\,147 \citep{Tasseroul2025}.
    \end{itemize}

\subsubsection{Composite SED}
Considering the above processes, we fit the observed SEDs with the following composite model 
\begin{align}
    S_{\rm{total}} &= A_{\rm sy}\,\left(\dfrac{\nu}{\nu_{0}}\right)^{\alpha_{\rm{sy}}} 
    \, e^{ - \dfrac{\nu_{\rm Razin}}{\nu} } \,  
    \left( \frac{1 - e^{-\tau (\nu)}}{\tau (\nu)} \right) \notag \\
    &\quad + 3.07 \times 10^4\,T_e\,\nu^2\,\Omega\,
    \left(1 - e^{-\tau(\nu)} \right),
    \label{eq:model-composite}
\end{align}
We fix the reference frequency at $\nu_0 = 1$~GHz. 
To reduce the number of free parameters, the electron temperature is fixed at $T_{\rm e} = 9000\,\rm K$, a typical value for ionized regions. We note that varying $T_{\rm e}$ within plausible bounds has a very small impact on the SED. 

In order to ensure that all possible solutions are explored and obtain reliable fits, we utilize a Bayesian inference approach that employs a Markov Chain Monte Carlo (MCMC) algorithm. In particular, we use the \texttt{EMCEE} sampler \citep{Foreman2013} implemented in the \texttt{Bilby} Python package \citep{Ashton2019}. We adopt uniform priors within a broad range of physically reasonable values for all fitted parameters. 
In particular, for $\alpha_{\rm sy}$ we use a prior range of $[-1.0, -0.5]$. However, in regions for which its posterior distribution peaks at the hard limit of $-0.5$, we fixed $\alpha_{\rm sy} = -0.5$. This value corresponds to the theoretical prediction of DSA in high Mach number adiabatic shocks \citep{Drury1983}, as expected in \ngc. 
Values harder than this ($\alpha_{\rm sy} > -0.5$) would imply that particles follow a power-law energy distribution with index $p < 2$, which is not favoured by particle-acceleration theory. 
The list of parameters and the corresponding explored ranges are given in Table \ref{tab:model-params}. 

The fitted SEDs are shown in Figure~\ref{fig:SED-fits}, and the corner plots presenting the complete details of the posterior distributions are provided in the Appendix \ref{sec:appendix-A}. The best fit parameters along with the chi-square ($\chi^2$) values and degree of freedom (d.o.f) are given in Table \ref{tab:bestfit-params}. 

\subsection{Parameter estimates and caveats}

The composite phenomenological model can reproduce well the observed SEDs and the behaviour of the spectral index (see Table~\ref{tab:WR7-spectralindex-sed}). 
To explain the observed turnover, $\nu_{\rm Razin}$ has to be $\sim$400~MHz.  Given its dependence on electron density and magnetic field (see Eq.~\ref{Eq:razin-fq}), this constrains the allowed values of $n_{\rm e}$ and $B$ (namely the ratio between these two quantities, as can be seen in the corner plots in Figures~\ref{fig:E1-4cornerplot} and \ref{fig:E5cornerplot}). Additionally, the electron density cannot be too high, as this would increase the free--free contribution and result in an incorrect slope at higher frequencies, inconsistent with the observed SED (which can also be seen in the corner plots in Figures~\ref{fig:E1-4cornerplot} and \ref{fig:E5cornerplot}). 
 Therefore, the turnover can only be reproduced if the magnetic field is sufficiently weak, given the range of electron densities allowed by the free--free emission constraints. %

The values of $n_{\rm e}$ that we derive are a factor of four less than the density estimated using the ionized mass and size of the whole complex reported by \citet{Cappa1999}; if we consider the entire bubble, it is an order of magnitude lower. However, these authors assumed that the emission arises entirely from thermal free--free radiation (which our analysis shows is not the case); hence, their ionized mass and number densities are likely overestimated.

Of interest is the low magnetic field strengths of $\sim0.5$--$4\,\mu$G derived by the model. 
In comparison, \citet{Prajapati2019} estimated the magnetic field strength in the synchrotron-emitting region of the bubble G2.4+1.4 to lie in the range of 120--770\,$\mu$G. This range corresponds to the assumed energy density ratio between NT particles and the magnetic field, ranging between 10 and 0.01. 
The synchrotron luminosity is proportional to the energy density in the magnetic fields and to the energy density in relativistic electrons. 
In this case, the weak magnetic field implies that an extremely large amount of energy must be injected into the relativistic particles to reproduce the observed synchrotron emission. Specifically, for a magnetic field of only a few $\mu\rm{G}$, using Eq.~\ref{eq:energy_density} we infer that the energy density in relativistic particles must be around $10^6$ times greater than the magnetic energy density. Such an extreme energy distribution may not be physically plausible, as it is unlikely to be sustained by the available stellar wind kinetic power.

The main caveat in our analysis is the assumption of a homogeneous emitting region, which is clearly an oversimplification for \ngc. Assuming a different density distribution would change the exact shape of the absorbed spectrum. For instance, in the case of FFA, for a homogeneous cloud, the spectral index is 2, while for a massive stellar wind where the density decreases with the square of the distance from the star, the index is close to 0.6 \citep{Wright1975}. This exemplifies that internal FFA can lead to a range of spectral indices in the observed spectrum.

A physically plausible scenario for \ngc could involve an inhomogeneous medium composed of dense, irregular substructures embedded within a lower-density environment. In such a configuration, FFA is not uniform across the region, as it depends on both the opacity and spatial distribution of the clumps. Radiation passing through denser regions can undergo significant absorption, while emission that travels through more diffuse interclump zones may remain largely unaffected. This selective filtering of emission, governed by the spatial variation in density and opacity, provides a more flexible framework for interpreting the observed SED features. A similar scenario of clumpy ionized regions is considered in different astrophysical environments \citep[e.g.,][]{Lacki2013,Conway2018,Mutie2025}. However, a quantitative assessment of this possibility would require an appropriate description of such complex structures, which can only be obtained through detailed magnetohydrodynamical simulations that are beyond the scope of this work. 

\section{Conclusions}
\label{sec:conclusion}

NT emission in massive stars is commonly detected in the colliding-wind region of binary systems under the \textit{wind--wind} interaction scenario \citep{DeBecker2013}.  
In contrast, only a few studies have investigated isolated, non-runaway WR stars with high wind kinetic power using low-frequency radio data. Among these, WR\,114 and WR\,142 \citep{Saha2023} are not associated with extended structures and no radio emission is detected from them, whereas WR\,102 \citep{Prajapati2019} and WR\,7 (this work) are surrounded by bubble-like structures and exhibit clear evidence of synchrotron emission arising from \textit{wind--ISM} interaction.
These results suggest that, for single stars, high wind kinetic power is a necessary but not sufficient condition for particle acceleration. 

The interaction between the stellar wind and the ambient ISM is crucial. Termination shocks in the stellar winds can provide the physical conditions required for efficient particle acceleration, leading to detectable synchrotron radiation.
This depends on the properties of the local ISM, such as density and temperature, together with stellar characteristics including how long the star remains in the WR phase, the stellar proper motion, mass of the swept-up material, and how efficiently wind energy is dissipated. However, it is unclear how these factors regulate the formation of strong shocks on parsec scales, underscoring the need for further observational and theoretical studies.


We investigated the WR bubble \ngc surrounding WR\,7 using uGMRT observations at low frequencies (250--500 MHz and 550--950 MHz), along with complementary archival datasets from GMRT, VLA, and ASKAP. 
The observed SEDs exhibit a steep negative slope above 1~GHz, a turnover below 1~GHz with a relatively flat slope between 400~MHz -- 1~GHz, and a steep positive slope at frequencies below 400~MHz. 
To extract physical information and understand the processes responsible for the spectral turnover, we modelled the observed SEDs using a composite model. This model includes contributions from both synchrotron and free--free emission, with two turnover mechanisms: internal FFA and the 
RT effect. We fitted the SEDs using the MCMC algorithm and estimated the electron number density and magnetic field strength.
Our analysis revealed synchrotron radiation from the bubble, characterized by spectral indices that are typically steeper than the canonical value of $-$0.5. The observed turnover is mainly due to the RT effect, with a small contribution from internal FFA. 
This is only the second detection of NT radiation from a stellar bubble associated with a single massive star, following the first discovery in G2.4+1.4. These findings confirm that \textit{wind--ISM} interactions in such bubbles can create the necessary conditions for accelerating particles to relativistic speeds, establishing that isolated massive stars can be considered potential sources of GCRs. 
Our analysis enables us to put tighter constraints on the electron number density, which is significantly less than the previous estimates \citep{Cappa1999}. The model also infers a very low magnetic field strength of a few $\mu$G. 
A potential caveat in our analysis is the assumption of a homogeneous emitting region. Based on the observed SEDs and derived model parameters, a more physically realistic scenario for \ngc involves an inhomogeneous medium, consisting of dense, irregular substructures embedded within a lower-density environment. 
Overall, it is crucial to investigate particle acceleration in a larger sample of WR bubbles by using data across a broad radio frequency range to gain insights into the necessary conditions that drive particle acceleration, underlying physical processes and various aspects of shock physics. 
With the upcoming Square Kilometre Array facility, significant breakthroughs are expected due to its very high sensitivity across different scales and broad frequency coverage below 1 GHz.

\begin{acknowledgements}
We thank the referee for valuable suggestions that helped us to
improve the manuscript. 
A.S acknowledges support from the National SKA Program of China (2025SKA0140100) and the National Natural Science Foundation of China (No. 12573025). 
A.S. and A.T. thank Dr. Dharam V. Lal from the National Centre for Radio Astrophysics for fruitful discussions. S.d.P. acknowledges support from ERC Advanced Grant 789410. 
R.S. and I.C.CH. acknowledge the support of the Department of Atomic Energy, Government of India, under project no. 12-R\&D-TFR-5.02-0700. 
This research is part of the PANTERA-Stars collaboration\footnote{\url{https://www.astro.uliege.be/~debecker/pantera}}, an initiative aimed at fostering research activities on the topic of particle acceleration associated with stellar sources. 
We thank the staff of the GMRT that made these observations possible. GMRT is run by the National Centre for Radio Astrophysics of the Tata Institute of Fundamental Research.
This research has made use of NASA's Astrophysics Data System Bibliographic Services. 
\end{acknowledgements}
%
\bibliographystyle{aa} 
\bibliography{reference} 
%
\begin{appendix}
\section{Posterior distribution: Corner plots}
\label{sec:appendix-A}
\begin{figure*}
    \centering
    \includegraphics[width=0.49\hsize]{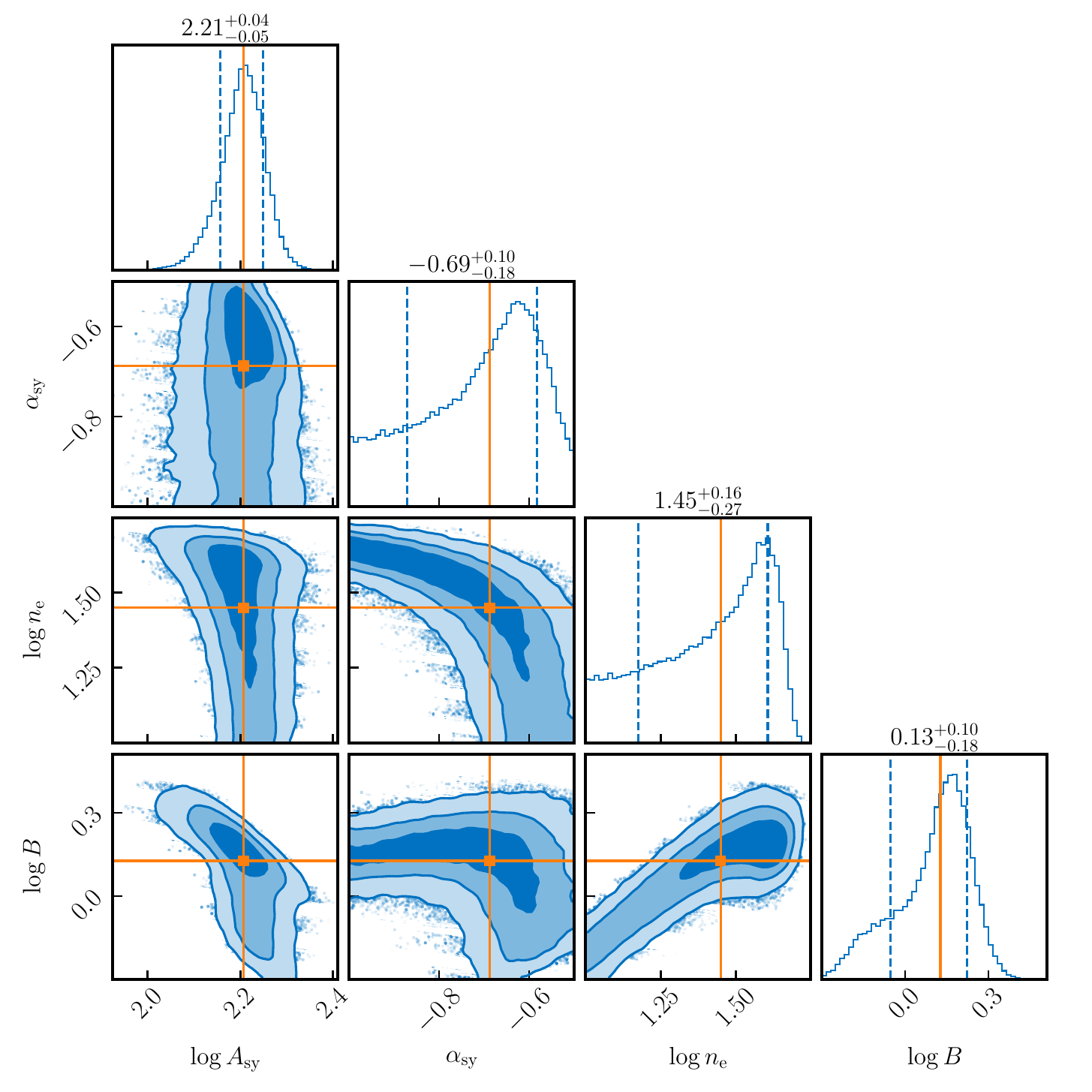}
    \includegraphics[width=0.49\hsize]{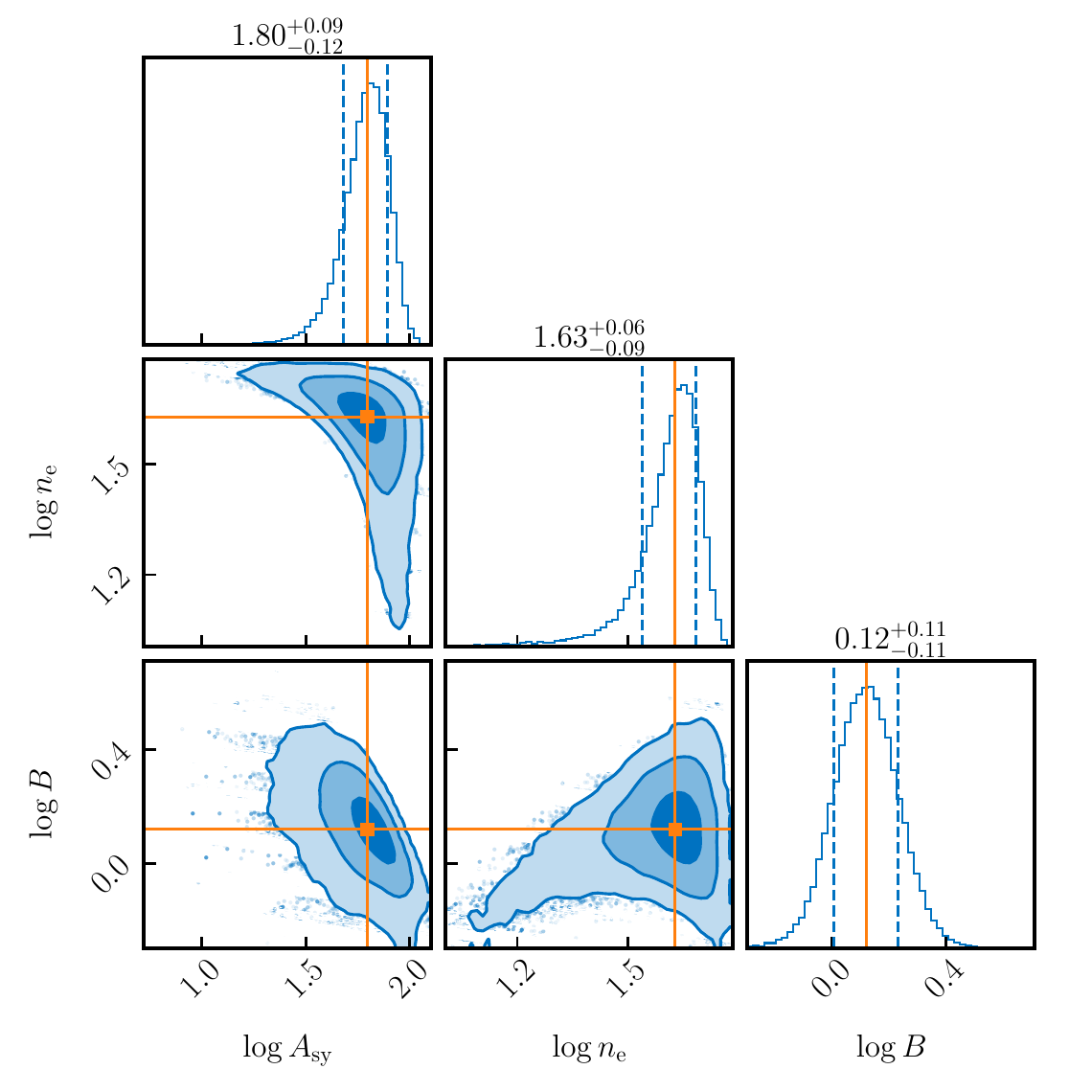}
    \includegraphics[width=0.49\hsize]{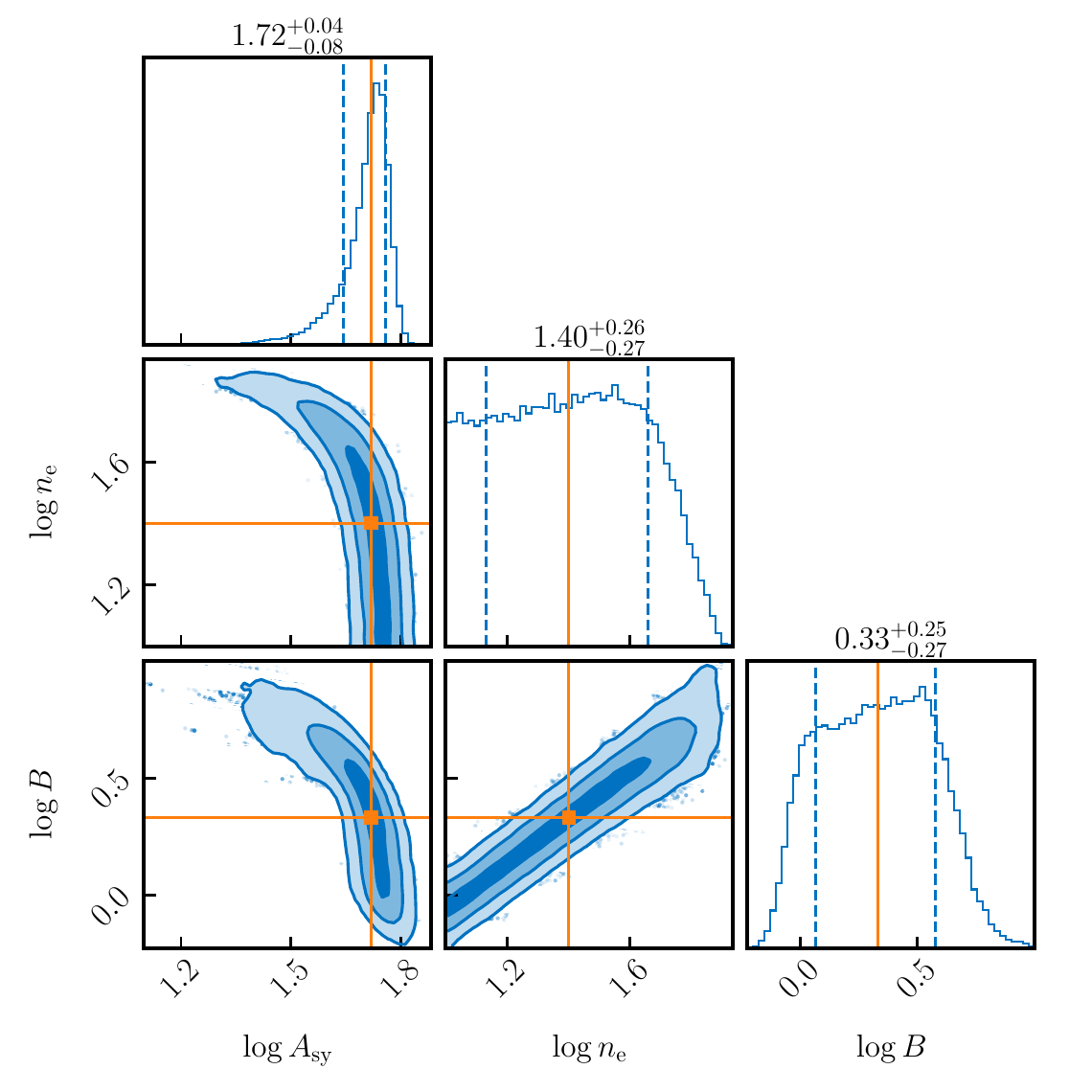}
    \includegraphics[width=0.49\hsize]{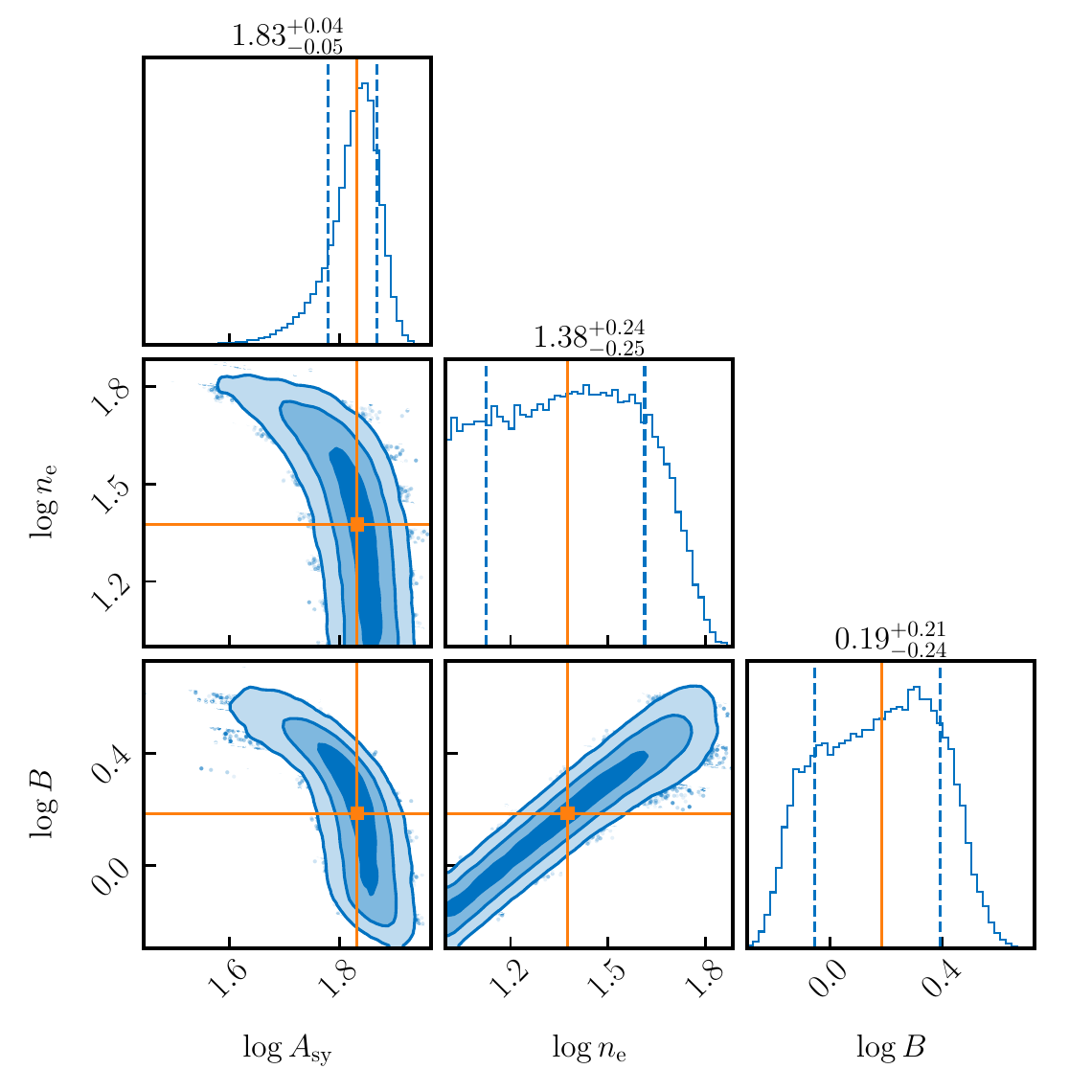}
    \caption{Cornerplot for the posteriors of the MCMC fitting for apertures, E1 (top left), E2 (top right), E3 (lower left) and E4 (lower right).}
    \label{fig:E1-4cornerplot}
\end{figure*}

\begin{figure}
    \centering
    \includegraphics[width=1.0\hsize]{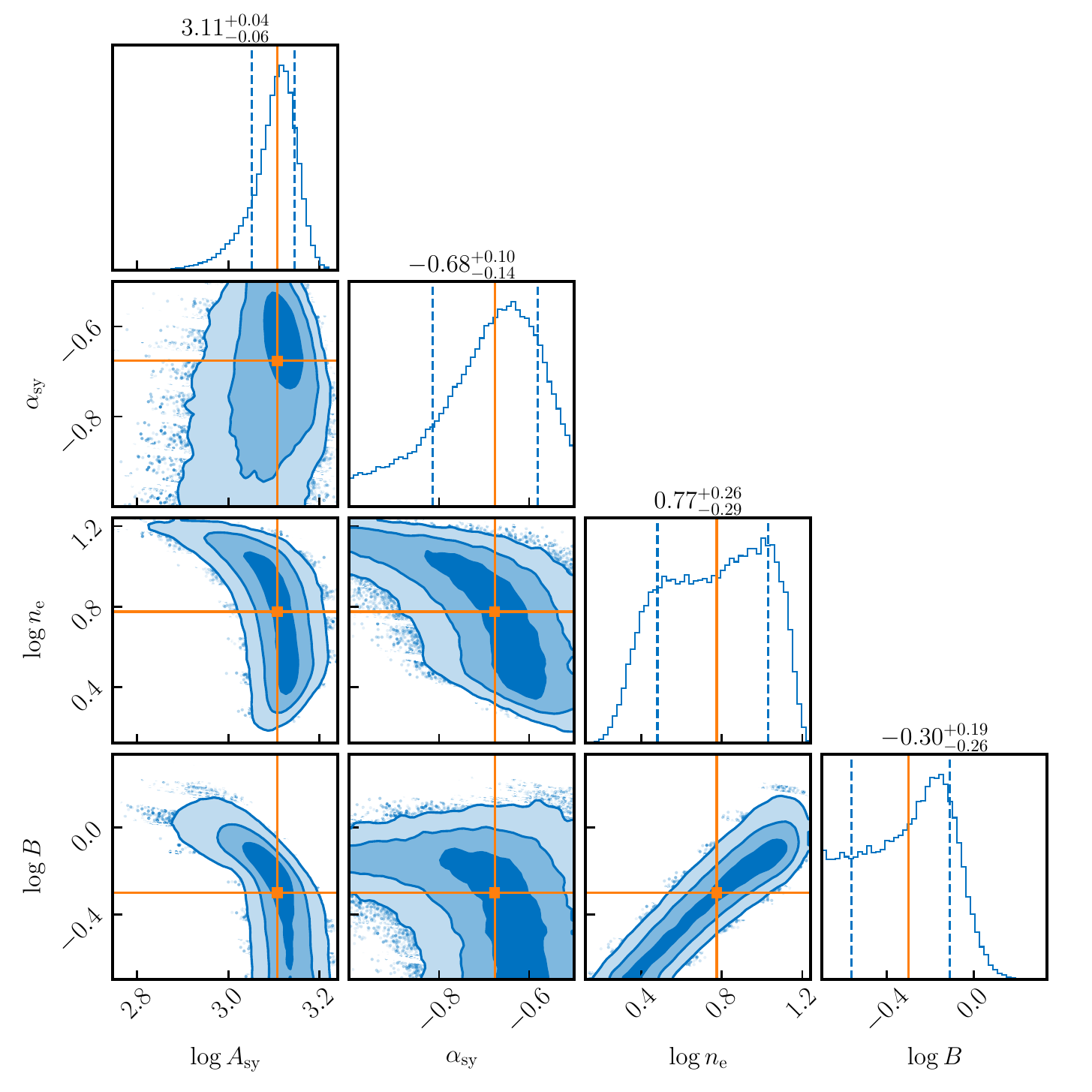}
    \caption{Cornerplot for the posteriors of the MCMC fitting for the central region of the bubble. }
    \label{fig:E5cornerplot}
\end{figure}

In Figs. ~\ref{fig:E1-4cornerplot} and ~\ref{fig:E5cornerplot} we present the corner plots showing the posterior distributions of the model parameters for all SED fits. The diagonal panels display the 1D distributions of each parameter, with the orange line indicating the best-fit value (median), and the dashed lines marking the 16th and 84th percentiles. The off-diagonal panels show the 2D joint distributions of parameter pairs, with contours representing the 1-, 2-, and 3-$\sigma$ confidence intervals. 

\end{appendix}

\end{document}